\documentclass[aps,preprint,prd,nofootinbib]{revtex4}

\usepackage[dvipsnames]{xcolor}
\usepackage{float}
\usepackage{ulem}
\usepackage[caption = false]{subfig}

\usepackage{graphicx}
\usepackage{epsfig}
\usepackage{epstopdf}
\usepackage[utf8]{inputenc}
\usepackage{amsmath}
\usepackage{amsfonts}
\usepackage{amssymb}
\usepackage{slashed}
\usepackage{booktabs}
\usepackage{multirow}
\usepackage{txfonts}
\allowdisplaybreaks

\newcommand{\nn}{\nonumber}

\begin{document}

\title{Threshold pion  electro- and photoproduction off nucleons in covariant chiral perturbation theory}

\author{Gustavo H. Guerrero Navarro}\email{gusguena@ific.uv.es}
\affiliation{Departamento de F\'{\i}sica Teorica and Instituto de Fisica Corpuscular (IFIC),
Centro Mixto UVEG-CSIC, Valencia E-46071, Spain}

\author{M. J. Vicente Vacas}\email{vicente@ific.uv.es}
\affiliation{Departamento de F\'{\i}sica Teorica and Instituto de Fisica Corpuscular (IFIC),
Centro Mixto UVEG-CSIC, Valencia E-46071, Spain}

\date{\today}

\begin{abstract}
Pion electro- and photoproduction off the nucleon close to threshold is  studied in covariant baryon chiral perturbation theory at O($p^3$)  in the extended-on-mass-shell scheme, with the explicit inclusion of the $\Delta(1232)$ resonance.  The relevant low energy constants are fixed by  fitting the available experimental data with the theoretical model. The inclusion of the $\Delta$ resonance as an explicit degree of freedom substantially improves the agreement with data and the convergence of the model.
 
\end{abstract}

\maketitle

\section{Introduction}

Since the first experiments in the early fifties \cite{Panofsky:1950gj}, little after  pion discovery,  electromagnetic pion production on nucleons has been a very important source of information about the nucleon-pion interaction, being also crucial in our knowledge of  several  baryonic resonances.
Here, we focus on this process near the threshold region, where there is a well founded theoretical framework to analyse it, namely, chiral perturbation theory (ChPT),  the low energy effective field theory based on the approximate chiral symmetry of quantum chromodynamics. 
Early theoretical efforts described electromagnetic pion production by means of  some low-energy-theorems (LET)~\cite{Kroll:1953vq} that were later extended using the partial conservation of the axial current (PCAC) and current algebra techniques~\cite{DeBaenst:1971hp,Vainshtein:1972ih}. 
While the LET results agreed well with the early  charged pion photoproduction 
data~\cite{Walker:1963zzb,Rossi:1973wf,Salomon:1983xn}, they couldn't explain  the neutral pion photoproduction on protons close to threshold. In particular, there was a clear disagreement for the $s$-wave electric dipole amplitude $E_{0+}$~\cite{Mazzucato:1986dz,Beck:1990da,Drechsel:1992pn}.
%
 These discrepancies were first solved in the framework of  ChPT~\cite{Bernard:1991rt}. At the lowest order, ChPT simply reproduces the LET results. However, higher order contributions  from chiral pion loops were found to lead to sizeable corrections 
and to an improvement of the agreement with the available data. 

Nevertheless, ChPT with baryons, such as it was used in Ref.~\cite{Bernard:1991rt}, was known to lack a systematic power-counting~\cite{Gasser:1987rb}. 
This consistency  problem was resolved by the heavy-baryon ChPT  (HBChPT) approach introduced in 
Refs.~\cite{Jenkins:1990jv,Jenkins:1991es} although at the expense of losing Lorentz covariance. 
Later,  a proper power-counting was also obtained in relativistic formulations of ChPT with the development of novel  schemes, like the infrared regularization (IR)~\cite{Becher:1999he} and the extended on-mass-shell (EOMS) formulation~\cite{Fuchs:2003qc}. 

Subsequently,  there has been extensive  work  using the  HBChPT framework. All the charge channels for pion electro- and photoproduction have been studied~\cite{Bernard:1992qa,Bernard:1992nc,Bernard:1992ys,Bernard:1993bq,Bernard:1994dt,Bernard:1994gm,Bernard:1995cj, Bernard:1996ti,Bernard:1996bi,Fearing:2000uy,Bernard:2001gz}  obtaining an overall good agreement with data and supporting the findings of Ref.~\cite{Bernard:1991rt}. However, the continuous  improvement of the quality and quantity of the experimental data unveiled some new problems. For instance,  data for electroproduction at low $Q^2$~\cite{Distler:1998ae,Merkel:2001qg,Merkel:2011cf}  were difficult to reproduce in HBChPT~\cite{Merkel:2011cf,Bernard:2007zu,Weis:2007kf}.  An   $O (q^4 )$ EOMS calculation~\cite{Hilt:2013fda} reached a good global agreement and fared better describing these low $Q^2$ data.

Other serious difficulties arose from the  $\pi^0$ photoproduction cross-section and polarized photon beam-asymmetry  measurements of the MAMI A2/CB-TAPS  experiment~\cite{Hornidge:2012ca}.  For this channel, both the covariant EOMS~\cite{Hilt:2013uf} as well as the HBChPT~\cite{FernandezRamirez:2012nw} approaches failed to reproduce the strong energy dependence of data even at  $O(q^4)$. They obtained a reasonable agreement with experiment only up to a mere $20~\text{MeV}$ above threshold. 
However, the chiral convergence and the  concordance with data of covariant  ChPT improved significantly with the  incorporation, as an explicit degree of freedom, of the lowest lying resonance  $\Delta(1232)$~\cite{Blin:2014rpa,Blin:2016itn}\footnote{
The possible importance of the $\Delta(1232)$ mechanisms was before suggested by Hemmert et al.~\cite{Hemmert:1996xg} and later in Refs.~\cite{Hornidge:2012ca,FernandezRamirez:2012nw}. 
The $\Delta$ role in  $\pi^0$ photoproduction has been also investigated in HBChPT  showing a rather important contribution~\cite{Cawthorne:2015orf}.
 }.
Indeed, it was well known phenomenologically that $\Delta(1232)$ mechanisms were  dominant in the $\pi^0$ photoproduction cross section,~(see, e.g.,  Ref. \cite{Ericson:1988gk}). 
Actually, the explicit inclusion of $\Delta(1232)$ leads to a  better agreement, and for a wider range of energies, at $O(q^3)$ than  other calculations,  even  at $O(q^4)$,  with only nucleon and pion degrees of freedom~\footnote{
The inclusion of $\Delta$ requires a modification of the power-counting scheme, due to the emergence of a new small parameter, $\delta=m_\Delta-m_N \approx 300 \; \text{MeV} $, in the $\Delta$ propagator for the scattering amplitudes.}.

Later, the same  approach of Refs.~\cite{Blin:2014rpa,Blin:2016itn},  EOMS  at $O(q^3)$ and with explicit  $\Delta$, was applied to investigate charged pion photoproduction in Ref~\cite{Navarro:2019iqj}.  It achieved  results consistent with data up to $E_\gamma=215 \; \text{MeV}$, about 70 MeV above threshold, for all observables.
Furthermore, many other processes have been investigated in this same framework.
For instance, this approach  obtained  a good overall reproduction of data and a fast convergence of the chiral series for Compton~\cite{Lensky:2009uv,Blin:2015era,Thurmann:2020mog} and $\pi N$ scattering~\cite{Alarcon:2012kn,Yao:2016vbz}, weak pion production~\cite{Yao:2018pzc,Yao:2019avf},
axial charges and form factors~\cite{Ledwig:2014rfa,Yao:2017fym}, electromagnetic form factors~\cite{Geng:2009hh, Blin:2017hez} or baryon masses~\cite{ Ren:2013dzt}.

Our aim in this work is to make a comprehensive analysis within the aforementioned framework of the electromagnetic pion production off nucleons. This study represents an extension of Ref.~\cite{Navarro:2019iqj} that considered only the photoproduction case. The inclusion of electroproduction allows for the exploration of the interaction of nucleons with virtual photons, and therefore to investigate some additional pieces of the chiral Lagrangian. This examination of the vector couplings of the nucleons might  reduce the large uncertainties that currently hinder our efforts to provide a theoretically well founded prediction of the neutrino induced pion production~\cite{Yao:2018pzc,Yao:2019avf}, a very important process in many of the neutrino experiments.

Furthermore, we will incorporate some recent  data for photoproduction of neutral \cite{Schumann:2015ypa} and charged pions \cite{Briscoe:2020qat}, and will  consider  explicit isospin breaking in the loop calculations. This latter point considerably improves the agreement with data at  low energies.

\section{Formalism and theoretical model}

\begin{figure}[ht]
\center
\includegraphics[scale=1.0]{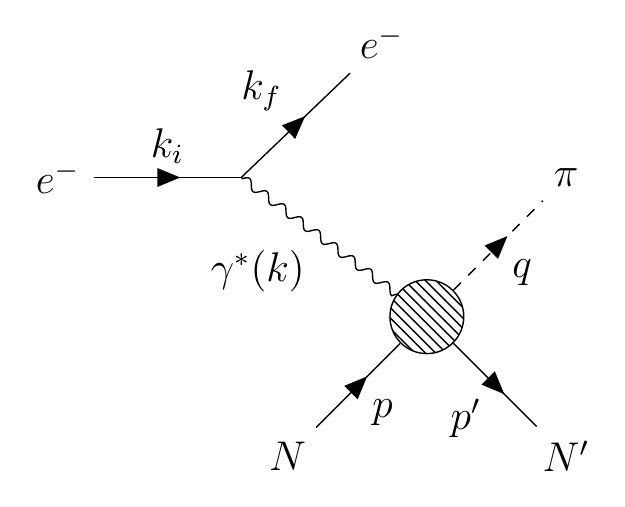}
\caption{Pion electroproduction on nucleons.}
\label{fig:process}
\end{figure}

We present here the basic formalism, our conventions and the studied observables for the pion electroproduction process depicted in Fig.~\ref{fig:process}. Other definitions useful for the analysis of the photoproduction channel can be found in Ref.~\cite{Navarro:2019iqj}.

\subsection{Kinematics}
The scattering amplitude $\mathcal{T}$ for the electroproduction of pions on nucleons, 
$e^{-} (k_i) + N(p) \rightarrow e^{-}(k_f) + N'(p') + \pi (q)$,
can be written in the one-photon exchange approximation as
\begin{equation}
\mathcal{T}=\frac{e}{k^2} \bar{u} (s_f,k_f) \gamma^\mu u (k_i, s_i) \mathcal{M}_\mu,
\label{eq:scattAmp}
\end{equation}
where 
\begin{equation}
\mathcal{M}_\mu =- i e \left\langle N', \pi  | J_\mu | N \right\rangle 
\label{eq:current}
\end{equation}
is the  electromagnetic  matrix element between the hadronic  states, which includes all the strong interaction dependence.  Here,   $k_{i,f} = \left(\mathcal{E}_{i,f}, \vec{k}_{i,f} \right)$ are the incoming and outgoing electron momenta, $s_i$ and $s_f$ are their spins, $k=k_i-k_f$ and $p$ are the incoming virtual-photon and nucleon momenta, while $q$ and $p'$ are the outgoing pion and nucleon momenta, respectively.

We also use the Mandelstam variables, defined as the invariants $s = (p + k)^2 = (p'+q)^2$, $u = (p-q)^2 = (p'-k)^2$, and $t =(p-p')^2 =(q-k)^2$.
They  satisfy the equation $s+t+u =  2 m_N^2 + M_\pi^2 -Q^2$,
where $m_N$ and $M_\pi$ are the nucleon and pion physical masses respectively and $Q^2=-k^2$.
 Moreover, we use the  angle between the outgoing pion and the incoming virtual-photon,  $\theta_\pi=\cos^{-1}(\hat{q}\cdot\hat{k})$, and $\phi_\pi$ defined as the angle between the scattering and the reaction planes given by $\hat{k}_i \times \hat{k}_f$ and $\hat{k} \times \hat{q}$ respectively.

For practical purposes, it is convenient to work in the final $\pi-N$ center of mass frame. There, we have $
\vec{p}^*= - \vec{k}^* $ for the initial nucleon and the virtual photon and 
$ \vec{p'}^*= - \vec{q}^*$ for final nucleon and pion. Also,
\begin{equation}
\begin{split}
E_\gamma^* =& \frac{1}{2 \sqrt{s}} \left( s- m_N^2 - Q^2 \right), \\
E_{\pi}^* =& \frac{1}{2 \sqrt{s}} \left( s + M_\pi^2 - m_N^2 \right),\\
E_{p}^* =& \frac{1}{2 \sqrt{s}} \left( s + m_N^2 + Q^2 \right), \\
E_{p'}^* =& \frac{1}{2 \sqrt{s}} \left( s + m_N^2 - M_\pi^2 \right),\\
|\vec{k}^*| =& \sqrt{{E_\gamma^*}^2 + Q^2} ,\\
|\vec{q}^*| =& \sqrt{{E_\pi^*}^2-M_\pi^2} ,\\
|\vec{p}^*| =& \sqrt{{E_p^*}^2 - m_N^2} ,\\
|\vec{p'}^*| =& \sqrt{{E_{p'}^*}^2 - m_N^2}. \\
\end{split}
\end{equation}
From here on,  except when explicitly otherwise indicated, all the four-vector components appearing in the formulas will correspond to the  $\pi - N$ center of mass frame, though omitting the asterisk symbol.

The scattering amplitude, $\mathcal{T}$,  can be written in terms of  the Chew-Goldberger-Low-Nambu (CGLN) basis, $\mathcal{F}_i$ \cite{Chew:1957tf,Dennery:1961zz}, 
\begin{equation}
\mathcal{T} = \epsilon^\mu \mathcal{M}_\mu =4 \pi \dfrac{W}{m_N} \chi_f^\dagger  \mathcal{F} \chi_i \; ,
\end{equation}
where $\epsilon^\mu =e/k^2 \bar{u}(p_f,s_f) \gamma^\mu u(p_i,s_i)$ is the virtual photon polarization vector, $\chi_i$ and $\chi_f$ denote the initial and final Pauli spinors, $W=\sqrt{s}$ is the invariant energy and the matrix $\mathcal{F}$ is written as
\begin{equation}
\begin{split}
\mathcal{F} =& i \vec{\tau} \cdot \vec{a}_\perp \mathcal{F}_1 + \frac{\vec{\tau} \cdot \vec{q} \vec{\tau} \cdot \vec{k} \times \vec{a}_\perp}{|\vec{q}||\vec{k}|} \mathcal{F}_2 + \frac{i \vec{\tau} \cdot \vec{k} \vec{q} \cdot \vec{a}_\perp}{|\vec{q}||\vec{k}|} \mathcal{F}_3  \\
& + \frac{i \vec{\tau} \cdot \vec{q} \vec{q} \cdot \vec{a}_\perp}{|\vec{q}|^2} \mathcal{F}_4 + \frac{i \vec{\tau} \cdot \vec{k} \vec{k} \cdot \vec{a}_\parallel}{|\vec{k}|^2} \mathcal{F}_5 + \frac{i \vec{\tau} \cdot \vec{q} \vec{k} \cdot \vec{a}_\parallel}{|\vec{q}||\vec{k}|} \mathcal{F}_6.
\end{split}
\label{eq:cglnF}
\end{equation}
Here, $\vec{\tau}=(\tau^1,\tau^2,\tau^3)$ are the Pauli matrices. The different contributions, transverse or parallel to the transferred momentum $\vec{k}$, are split with the help of the $\vec{a}_{\perp}$ and $\vec{a}_{\parallel}$ vector components. The four-vector $a^\mu$ is defined such that its time component is zero,  by~\cite{Amaldi:1979vh}
\begin{align}
a^\mu = \epsilon^\mu - k^\mu \frac{\epsilon_0}{E_\gamma} = \epsilon^\mu - k^\mu \frac{\vec{k} \cdot \vec{\epsilon}}{E_\gamma^2},
\end{align}
where the Lorentz condition, $k_\mu \epsilon^\mu=0$, has been used and
\begin{align}
\vec{a} =& \vec{a}_\parallel + \vec{a}_\perp, \\
\vec{a}_\parallel = & \vec{a} \cdot \hat{k} \hat{k} = \frac{k^2}{E_\gamma^2} \vec{\epsilon} \cdot \hat{k} \hat{k}, \\
\vec{a}_\perp = & \vec{a}- \vec{a}_\parallel= \vec{\epsilon} - \vec{\epsilon} \cdot \hat{k} \hat{k}=\vec{\epsilon}_\perp.
\end{align}

\subsection{Observables}

For an electroproduction experiment, the differential cross section can be written as~\cite{Drechsel:1992pn}
\begin{equation}
\frac{d \sigma}{d \Omega_f d \mathcal{E}_f d \Omega_\pi}= \Gamma \frac{d \sigma_v}{d \Omega_\pi},
\end{equation}
where the flux of the virtual photon field is 
\begin{equation}
\Gamma = \frac{\alpha}{2 \pi^2} \frac{\mathcal{E}_f}{\mathcal{E}_i} \frac{k_\gamma^{lab}}{Q^2} \frac{1}{1-\varepsilon},
\end{equation}
$k_\gamma^{lab}=(W^2-m^2_N)/2m_N$ is the equivalent photon energy in the laboratory frame, $\alpha=e^2/4\pi \sim 1/137$,
\begin{equation}
\varepsilon = \left( 1 + \frac{2 |\vec{k}|^2}{Q^2} \tan^2 \frac{\Theta_e}{2} \right)^{-1}
\end{equation}
is the transverse polarization of the virtual photon~\cite{Hilt:2011imm, Knochlein:1995qz} with $\Theta_e$ the electron scattering angle. The parameter $\varepsilon$ is an invariant under collinear transformations, {\it i.e.}, $\vec{k}$ and $\Theta_e$ may be both expressed  in the lab. or in the c.m. frame.
The virtual photon differential cross section, $d \sigma_v/d \Omega_\pi$, for an unpolarized target and without recoil polarization can be cast in the form~\cite{Drechsel:1992pn,Hilt:2013fda}\footnote{A slightly different notation in terms of the longitudinal polarization, $\varepsilon_L = (Q^2 / E_\gamma^2) \varepsilon$, is used in Ref.~\cite{Drechsel:1992pn}.}
\begin{equation}
\begin{split}
\frac{d \sigma_v}{d \Omega_\pi} =& \frac{d \sigma_T}{d \Omega_\pi} + \varepsilon \frac{d \sigma_L}{d \Omega_\pi} + \sqrt{2 \varepsilon (1+ \varepsilon)} \frac{d \sigma_{LT}}{d \Omega_\pi} \cos \phi_\pi + \varepsilon \frac{d \sigma_{TT}}{d \Omega_\pi} \cos 2 \phi_\pi \\
& + h \sqrt{2 \varepsilon (1-\varepsilon)} \frac{d \sigma_{LT'}}{d \Omega_\pi} \sin \phi_\pi\;,
\end{split}
\label{eq:virtualsigma}
\end{equation}
where $h$ indicates the electron helicity, the subscripts refer to the transverse, $T$, and longitudinal, $L$, components. The two first terms are independent of the azimuthal angle $\phi_\pi$. The $\phi_\pi$ dependence is explicit and is decomposed in the $LT$ and $LT'$ pieces, related to the transverse-longitudinal interference, and the transverse-transverse term, $TT$,  which is proportional to $\sin 2\phi_\pi$. The different components of  Eq. (\ref{eq:virtualsigma}), can be given in terms of the diverse longitudinal and transverse response functions~\cite{Hilt:2013fda},
\begin{equation}
\begin{split}
\frac{d \sigma_T}{d \Omega_\pi} = & \rho_0 R_T, \\
\frac{d \sigma_L}{d \Omega_\pi} = & \rho_0 \dfrac{Q^2}{E_\gamma^2} R_L,\\
\frac{d \sigma_{LT}}{d \Omega_\pi} = & \rho_0 \dfrac{Q}{|E_\gamma|} R_{LT},\\
\frac{d \sigma_{TT}}{d \Omega_\pi} = & \rho_0 R_{TT},\\
\frac{d \sigma_{LT'}}{d \Omega_\pi} = & \rho_0 \dfrac{Q}{|E_\gamma|} R_{LT'}.
\end{split}
\label{eq:observables}
\end{equation}
Here, the phase space factor $\rho_0=|\vec{q}|/k_\gamma^{cm} $ with $k_\gamma^{cm}=k_\gamma^{lab} m_N/W$.  Finally, the response functions, in terms of the CGLN basis, are given by~\cite{Knochlein:1995qz}
\begin{equation}
\begin{split}
R_T =& |\mathcal{F}_1|^2 + |\mathcal{F}_2|^2 + \frac{\sin^2 \theta_\pi}{2} \left( |\mathcal{F}_3|^2 + |\mathcal{F}_4|^2\right) \\
&+ \mathfrak{R}e \left\{ \sin^2 \theta_\pi \left( \mathcal{F}_2^* \mathcal{F}_3 + \mathcal{F}_1^* \mathcal{F}_4 + \cos \theta_\pi \mathcal{F}_3^* \mathcal{F}_4 \right) \right. \\
& \left. - 2 \cos \theta_\pi \mathcal{F}_1^* \mathcal{F}_2 \right\} ,\\
R_L =& \mathfrak{R}e \left\{ |\mathcal{F}_5|^5 + |\mathcal{F}_6|^2 + 2 \cos \theta_\pi \mathcal{F}_5^* \mathcal{F}_6 \right\} ,\\
R_{LT} =& \sin \theta_\pi \mathfrak{R}e \left\{ -\mathcal{F}_2^* \mathcal{F}_5 - \mathcal{F}_3^* \mathcal{F}_5 - \mathcal{F}_1^* \mathcal{F}_6 - \mathcal{F}_4^* \mathcal{F}_6 \right. \\
& \left. - \cos \theta_\pi \left( \mathcal{F}_4^* \mathcal{F}_5 + \mathcal{F}_3^* \mathcal{F}_6 \right) \right\} ,\\
R_{TT} =& \frac{1}{2} \sin^2 \theta_\pi \left\{ |\mathcal{F}_3|^2 + |\mathcal{F}_4|^2 \right\} \\
& + \sin^2 \theta_\pi \mathfrak{R}e \left\{ \mathcal{F}^*_2 \mathcal{F}_3 + \mathcal{F}^*_1 \mathcal{F}_4 + \cos \theta_\pi \mathcal{F}^*_3 \mathcal{F}_4 \right\} ,\\
R_{LT'} =& - \sin \theta_\pi \mathfrak{I}m \left\{ \mathcal{F}^*_2 \mathcal{F}_5 +\mathcal{F}^*_3 \mathcal{F}_5 + \mathcal{F}^*_1 \mathcal{F}_6 + \mathcal{F}^*_4 \mathcal{F}_6 \right. \\
& \left. + \cos \theta_\pi \left( \mathcal{F}^*_4 \mathcal{F}_5 + \mathcal{F}^*_3 \mathcal{F}_6 \right) \right\}.
\end{split}
\label{eq:response_func}
\end{equation}
Most of the experimental data correspond to some of the terms appearing in Eq.~(\ref{eq:virtualsigma}). 
Additionally, an observable proportional to $d \sigma_{LT'}/d \Omega_\pi$ has been  measured~\cite{Weis:2007kf},
\begin{align}
A_{L T'} = \frac{\sigma^+ - \sigma^-}{\sigma^+ + \sigma^-} = \frac{\sqrt{2 \varepsilon(1- \varepsilon)} d \sigma_{LT'}}{d \sigma_T + \varepsilon d\sigma_L - \varepsilon d \sigma_{TT}},
\label{eq:obs_asym}
\end{align}
where $\sigma^+$ and $\sigma^-$ are the differential cross sections for $\phi_\pi = 90^\circ$ with beam polarization parallel and antiparallel to the beam direction, respectively.

\subsection{Theoretical model for electroproduction}

We analyse the electromagnetic  pion production process close to threshold using ChPT  up through order $\mathcal{O}(p^3)$. 
Here, $p$ is  a small parameter controlling the chiral expansion such as  the pion mass,  $M$, or  $q(k)$ the  pion(photon) momentum.
In particular, we consider the low order chiral Lagrangian terms for nucleon,  $\Delta(1232)$, pions and photons. For our calculation  the following set of Lagrangian pieces is required
\begin{equation}
\mathcal{L}_{\text{eff}} = \sum_{i=1}^2 \mathcal{L}_{\pi \pi}^{(2i)} + \sum_{j=1}^3 \mathcal{L}_N^{(j)} + \mathcal{L}_{\Delta N \pi}^{(1)} + \mathcal{L}_{\Delta N \gamma}^{(2)}\,.
\end{equation}
The superscripts indicate the chiral order. 
In the evaluation of  the  hadron electromagnetic current for the process $\gamma^* N \rightarrow \pi N'$,  $\mathcal{M}^\mu$,
the chiral order for a Feynman diagram with  $L$ loops, $V^{(k)}$ vertices of order $k$,  $n_\pi$ internal pions, $n_N$ nucleon and $n_\Delta$ $\Delta(1232)$ propagators,
  is given by 
\begin{align}
D=4 L \sum_{k=1}^{\infty} k V^{k}-2 n_\pi - n_N -\frac{1}{2}n_\Delta.
\label{eq:delta-counting}
\end{align}
Here, keeping consistency with our previous work on photoproduction~\cite{Navarro:2019iqj}, we use the $\delta$ power counting rule~\cite{Pascalutsa:2002pi}
for which a $\Delta$-propagator  contributes at $\mathcal{O}(p^{1/2})$ in the chiral expansions~\footnote{The $\delta$ counting  is appropriate at low energies. There, we have the energy $\omega\sim m_\pi\ll \delta \ll 4\pi F_{\pi} $, and to keep this hierarchy one takes $\delta^2\sim m_{\pi}$.}. 

\subsubsection{Nucleon and pion degrees of freedom}
The relevant Lagrangian terms in the  mesonic sector  are~\cite{Gasser:1987rb} 
\begin{align}
\mathcal{L}_{\pi\pi}^{(2)} &= \frac{F^2}{4} \text{Tr} \left[ \nabla^\mu U \left( \nabla_\mu U \right)^\dagger  + \chi U^\dagger + U \chi^\dagger \right],
\label{eq:LOLagpipi}\\
\mathcal{L}_{\pi\pi}^{GSS(4)} &=\frac{l_3+l_4}{16} \text{Tr}\left[ \chi U^\dagger + U \chi^\dagger \right]^2  +  \frac{l_4}{8} \text{Tr} \left[ \nabla_\mu U \left[ \nabla^\mu U \right]^\dagger \right] \text{Tr} \left[ \chi U^\dagger + U \chi^\dagger \right] \nonumber \\
&+ i \frac{l_6}{2} \text{Tr} \left[ F_{R \mu \nu} \nabla^\mu U \left( \nabla^\nu U \right)^\dagger + F_{L \mu \nu} \left( \nabla^\mu U \right)^\dagger \nabla^\nu U \right] + \cdots\,,
\label{eq:LagPi4}
\end{align}
where the ellipsis indicates terms that are not needed in the calculation. Pions are represented by the matrix function
\begin{align}
U= \exp \left[ i \frac{\vec{\tau} \cdot \vec{\pi}}{F} \right], \hspace{0.5cm} \vec{\tau} \cdot \vec{\pi}=
\begin{pmatrix}
    \pi^0            &  \sqrt{2} \pi^+\\
    \sqrt{2} \pi^-  & -\pi^0
\end{pmatrix},
\end{align}
with $\pi_i$ the cartesian pion fields, $F$ is the chiral limit of the pion decay constant  $F_\pi$, Tr$[...]$ indicates the trace of the resulting matrix in the isospin space, $\nabla_\mu U = \partial_\mu U -i r_\mu U + i U l_\mu $ is the covariant derivative for the pion,  $l_\mu$ and $r_\mu$ are  left- and right-handed external fields. For the electromagnetic case $r_\mu = l_\mu = e Q A_\mu$ with $e$ the electron charge, $Q=\frac{1}2(\tau_3 + {\mathbf{1}}_{2 \times 2})$ the charge matrix and $A_\mu$ the photon field. Moreover, the  matrix $\chi=M^2 \mathbf{1}_{2 \times 2}$ accounts for the explicit chiral symmetry breaking that leads to the pion mass. Finally,
\begin{align}
F^{\pm}_{\mu \nu} &= u^\dagger F_{R \mu \nu} u \pm u F_{L \mu \nu} u^\dagger, \\
F_{\mu \nu}&= e Q (\partial_\mu A_\nu - \partial_\nu A_\mu), \\
F_{R \mu \nu} &= F_{L \mu \nu} = F_{\mu \nu}.
\end{align}

For the nucleonic sector, the contributing Lagrangian terms are given by~\cite{Fettes:2000gb}
\begin{align}
\mathcal{L}_N^{(1)} &= \bar{N} \left( i \slashed D - m + \frac{g}{2} \slashed u \gamma_5 \right) N  ,
\label{LN1} \\
\mathcal{L}_{N}^{(2)}  &= \bar{N} \left( c_1 \text{Tr}\left[ \chi_{+} \right]+  \frac{c_6}{8m_N} F_{\mu\nu}^{+}\sigma^{\mu\nu}  + \frac{c_7}{8m_N} \text{Tr} \left[ F_{\mu\nu}^{+} \right] \sigma^{\mu\nu} \right) N +\cdots,
\label{LN2} \\
\mathcal{L}_N^{(3)} &= d_6 \bar{N} \left( \frac{1}{2m_N}i[D^{\mu},\widetilde{F}_{\mu\nu}^{+}]D^{\nu}+{\rm H.c.} \right) N  + d_7 \bar{N} \left( \frac{1}{2m_N}i[D^{\mu},\text{Tr} \left[ F_{\mu\nu}^{+}\right] ]D^{\nu}+{\rm H.c.} \right) N  \nonumber \\
&+d_8 \bar{N} \left( \frac{1}{2m_N}i\epsilon
^{\mu\nu\alpha\beta}\text{Tr} \left[ \widetilde{F}_{\mu\nu}^{+}u_{\alpha} \right] D_{\beta}+{\rm H.c.} \right) N + d_9 \bar{N} \left( \frac{1}{2m_N}i\epsilon^{\mu\nu\alpha\beta}\text{Tr} \left[  F_{\mu\nu}^{+}\right] u_{\alpha}D_{\beta}+{\rm H.c.} \right) N \nonumber \\
&+ d_{16} \bar{N} \left( \frac{1}{2}\gamma^{\mu}\gamma_{5} \text{Tr} \left[ \chi_{+}\right] u_{\mu} \right) N + d_{18} \bar{N} \left( \frac{1}{2}\,i\gamma^{\mu}\gamma_{5}[D_{\mu},\chi_{-}] \right) N \nonumber \\
&+ d_{20} \bar{N} \left( -\frac{1}{8m_N^{2}} i\gamma^{\mu}\gamma_{5}[\widetilde{F}_{\mu\nu}^{+},u_{\lambda}]D^{\lambda\nu}+{\rm H.c.} \right) N  \nonumber \\
&+ d_{21} \bar{N} \left( \frac{1}{2}i\gamma^{\mu}\gamma_{5} [\widetilde{F}_{\mu\nu}^{+},u^{\nu}] \right) N + d_{22} \bar{N} \left( \frac{1}{2}\,\gamma^{\mu}\gamma_{5}[D^{\nu},F_{\mu\nu}^{-}] \right) N + \cdots \,,
\end{align}
where $N=\left( p, n\right)^\text{T}$ is the nucleon isospin doublet with mass $m$ and axial charge $g$, both in the chiral limit.  The covariant derivative operator for the nucleon field is given by $D_\mu =\partial_\mu + \Gamma_\mu$ with $\Gamma_\mu=\frac{1}{2} [u^\dagger, \partial_\mu u] - \frac{i}{2} u^\dagger r_\mu u - \frac{i}{2} u l_\mu u^\dagger$. Moreover,
\begin{align}
u_\mu &= i u^\dagger \nabla_\mu U u^\dagger, \\
u&= U^{1/2}, \\
\chi_{\pm} &= M^2 \left( U^\dagger \pm U \right), \\
\sigma^{\mu \nu} &= \frac{i}{2} \left[ \gamma^\mu ,\gamma^\nu \right] ,\\
\widetilde{F}_{\mu\nu}^{+} &= {F}_{\mu\nu}^{+} - \frac{1}{2} \text{Tr}\left[ {F}_{\mu\nu}^{+} \right],\\
D_{\mu \nu}&= \left\lbrace D_\mu , D_\nu \right\rbrace .
\end{align}
Considering the hitherto presented terms, with only nucleon, pion and photon degrees of freedom, we generate the tree level contributions for the $\gamma^* N \rightarrow \pi N'$ reaction represented by the Feynman diagrams in  Fig.~\ref{fig:tree}. 
The explicit  expressions for the  associated amplitudes  are given  in the Appendix,  Sec.~\ref{app:amplitudes}.
\begin{figure}
    \centering
    \includegraphics{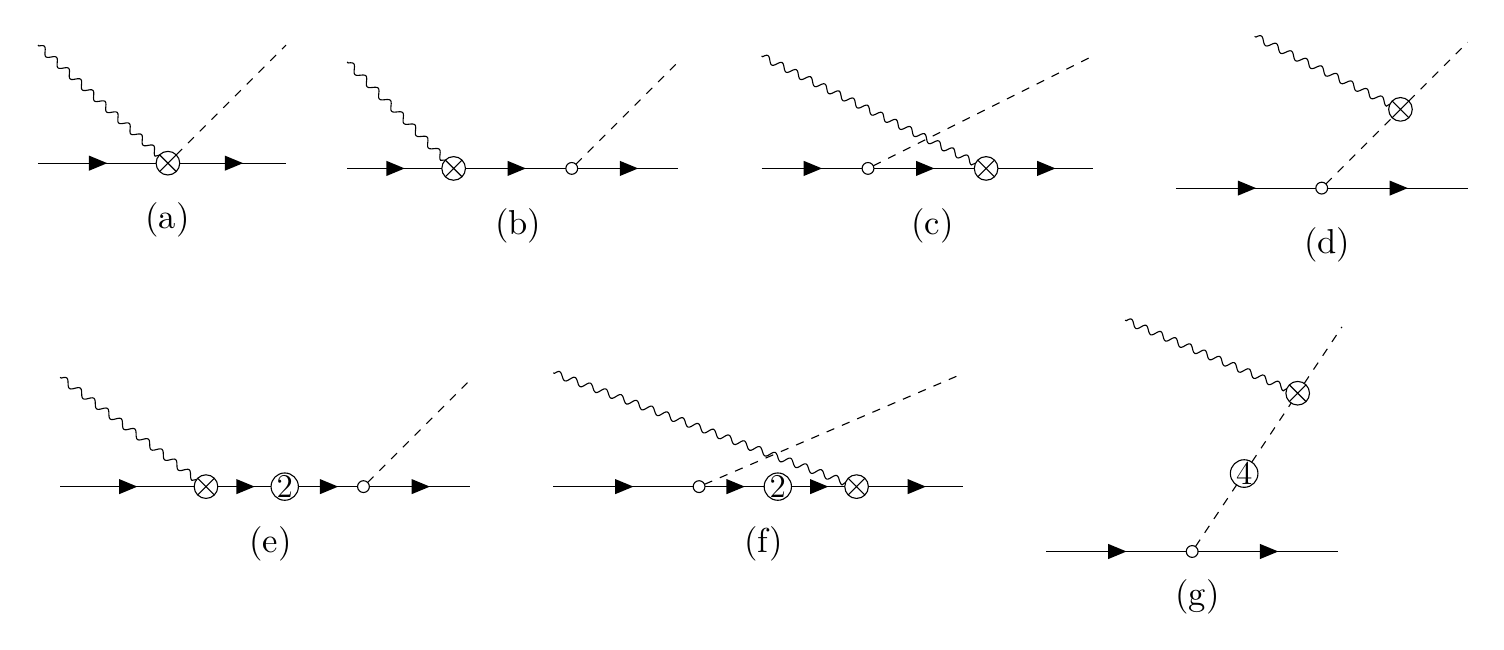}
    \caption{Tree level  diagrams for the pion electroproduction process: (a) contact term, (b)-(d) including propagators in the chiral limit and (e)-(g) including a mass correction in the propagator. Numbers inside the  circles indicate the chiral order of the vertex and crossed circles stand for vertices with an incoming photon.}
    \label{fig:tree}
\end{figure}

\begin{figure}
    \centering
    \includegraphics{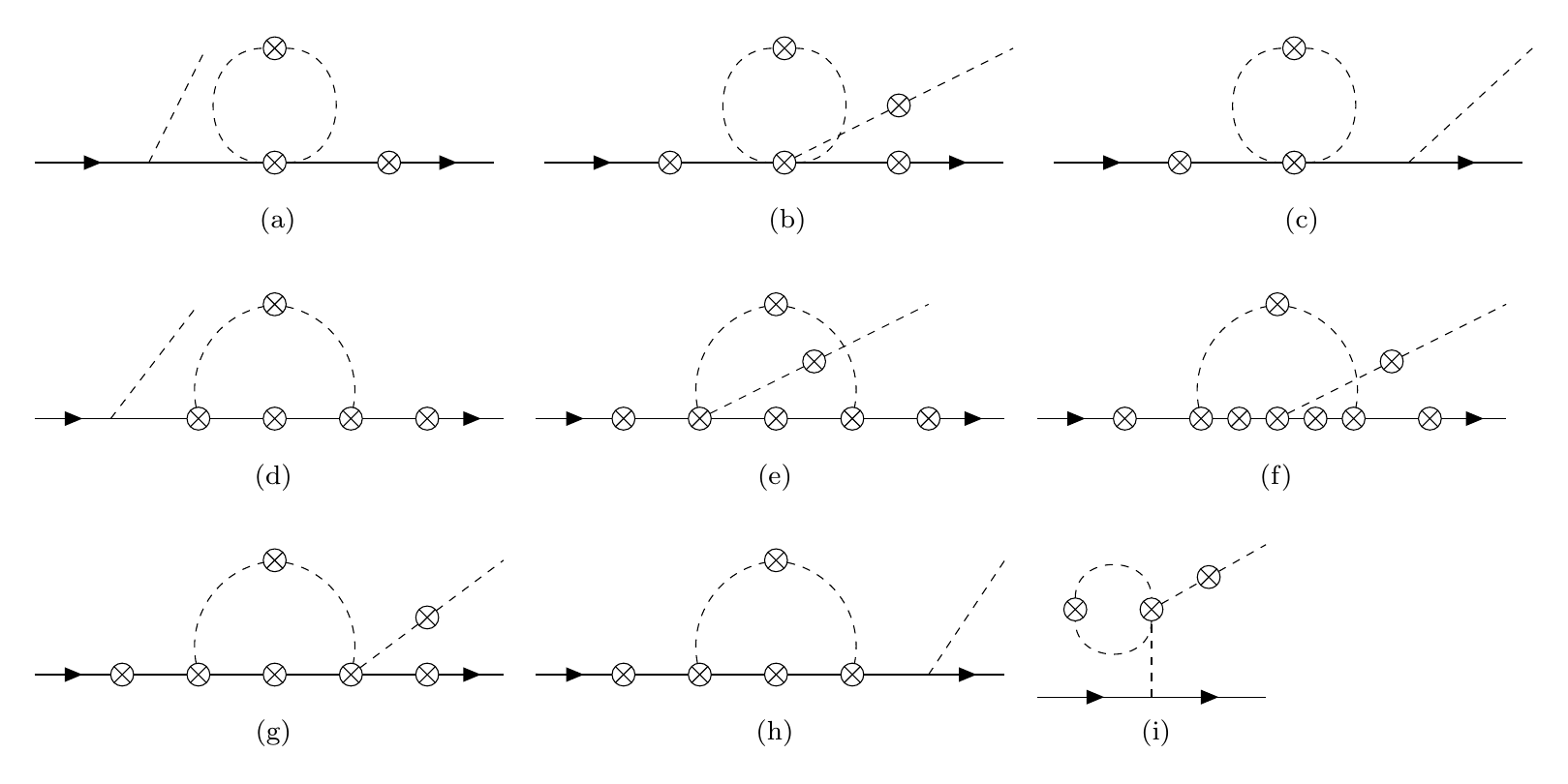}
    \caption{One loop topologies for pion electroproduction from which Feynman diagrams are generated. Solid lines are nucleons, dashed lines are pions. Crossed circles indicate where a virtual photon can be inserted. The topologies that lead to loop corrections in the external pion and nucleon legs are not shown because they are taken into account by the wave function renormalization.
}
    \label{fig:oneloop}
\end{figure}

There are many  one-loop diagrams contributing at $\mathcal{O}(p^3)$ . The generating  topologies are depicted in Fig.~\ref{fig:oneloop}.
 The  amplitudes have been computed with the help of Mathematica and the FeynCalc package \cite{Mertig:1990an,Shtabovenko:2016sxi}. The explicit expressions can be obtained from the authors upon request.
The UV divergences from the one-loop amplitudes, are  subtracted in the modified minimal subtraction scheme ($\overline{\text{MS}}$-1 or $\widetilde{\text{MS}}$)\footnote{In this scheme, multiples of  $\gamma_E-1/\epsilon_{UV} -\log (4\pi)-1$ are subtracted, where $\epsilon_{UV}=(4-d)/2$, with $d$ the space-time dimension, and $\gamma_E$ the Euler-Mascheroni constant.}. We  take the renormalization scale $\mu= m_N$, the nucleon mass.

As mentioned before, loop diagrams with internal nucleon propagators can give rise to analytical terms of orders below the nominal one, Eq.~(\ref{eq:delta-counting}).  We follow the EOMS procedure to restore the power counting. Namely, the power counting breaking terms (PCBT) are proportional to lower order tree-level amplitudes and in consequence can be subtracted  by finite shifts of the appropriate LECs, in our case those at $\mathcal{O}(p^1)$ and $\mathcal{O}(p^2)$.  Thus, after the UV renormalization, we apply the following substitution
\begin{align}
X= \widetilde{X} + \frac{m \widetilde{\beta}_X}{16 \pi^2 F^2}\,,
\end{align}
where $X \in \left\{ m , g, c_1, c_6, c_7 \right\}$ are the  shifted LECs,  $\widetilde{X}$ the corresponding EOMS parameters, and $\widetilde{\beta}_X$  are the  proportionality constants needed to generate the  terms that cancel the PCBT. Their values are shown in the  Appendix, Sec.~\ref{app:eomsbeta}.

Additionally, there are diagrams with loop insertions in the external legs that are not shown in Fig.~\ref{fig:oneloop}.
Their contribution is considered systematically via the Lehmann-Symanzik-Zimmermann reduction 
formula~\cite{Lehmann:1954rq},
\begin{align}
\mathcal{M}^\mu = \sqrt{\mathcal{Z}_\pi} \mathcal{Z}_N \hat{\mathcal{M}}^\mu ,
\end{align}
where $\hat{\mathcal{M}}^\mu$ is the so-called amputated amplitude as obtained from Figs.~\ref{fig:tree}-\ref{fig:oneloop}, and the missing pieces are encoded  in the wave function renormalization for the nucleons $\mathcal{Z}_N$ and pion $\mathcal{Z}_\pi$. Their explicit expressions are given in the Appendix, Sec.~\ref{app:wfr}.

\subsubsection{Contribution of the  $\Delta(1232)$ resonance}

\begin{figure}
    \centering
    \includegraphics{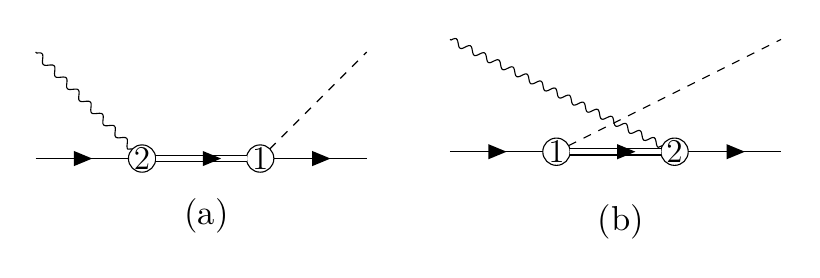}
    \caption{ Feynman diagrams including the contribution of the $\Delta$ resonance to pion electroproduction.  Numbers indicate the chiral order of the vertex.}
    \label{fig:delta}
\end{figure}

The only mechanisms involving the $\Delta$ resonance and contributing  to $\gamma^* N \rightarrow \pi N'$ up to $\mathcal{O}(p^3)$ are shown in Fig.~\ref{fig:delta}. Loop diagrams with a $\Delta$ propagator start at $\mathcal{O}(p^{7/2})$, beyond our current scope.
The relevant  Lagrangian terms are \cite{Blin:2016itn,Pascalutsa:2007yg}
\begin{equation}
\mathcal{L}_{\Delta N \pi}^{(1)}=\frac{i h_A}{ 2 F m_\Delta} \bar{N} T^a \gamma^{\mu \nu \lambda}  (\partial_\mu \Delta_\nu) \partial_\lambda \pi^a + \text{h.c.},
\end{equation}
\begin{equation}
\mathcal{L}_{\Delta N \gamma}^{(2)}=\frac{3 i e g_M}{2m(m+m_\Delta)} \bar{N} T^3 (\partial_\mu \Delta_\nu) \widetilde{f}^{\mu \nu} + \text{h.c.},
\end{equation}
with $h_A$ that can be fixed from the strong $\Delta\rightarrow\pi N$ decay, and  $g_M$ from the electromagnetic one, $\Delta\rightarrow\gamma N$. Also, $ \gamma^{\mu \nu \lambda}= \frac{1}{4} \left \{ \left[ \gamma^\mu , \gamma^\nu \right], \gamma^\lambda \right \}$ and
$\widetilde{f}^{\mu \nu}=\frac{1}{2} \epsilon^{\mu \nu \alpha \beta} 
( \partial_\alpha A_\beta - \partial_\beta A_\alpha)$. The $\Delta(1232)$ isospin multiplet is given by
 $\Delta_\nu =(\Delta^{++}_\nu, \Delta^+_\nu,\Delta^0_\nu,\Delta^-_\nu)^T $ and the isospin transition matrices $T^a$ can be found in Ref.~\cite{Pascalutsa:2006up}.

\subsubsection{Isospin symmetry treatment}
As it is obvious from our choice of the Lagrangian, the vertices are calculated  in the isospin symmetric limit ($m_u=m_d$). However, the physical masses of pions and nucleons are used in the evaluation of the loops. Formally, in our  $\mathcal{O}(p^3)$ calculation, this amounts to a higher order correction. Nonetheless, it allows to properly  reproduce the cusp, due to the different thresholds for the two charge channels,  clearly visible in the $E_{0+}$ multipole for the $\gamma p\rightarrow \pi^0 p$ reaction~\cite{Bernard:1993bq}. In general, it should lead to some visible changes very close to threshold,  where the isospin mass splittings could be relevant, while producing only small numerical changes at higher energies.


\subsection{Low-energy-constants and fitting procedure}
\begin{table}
\caption{ Values of the LECs determined from other processes.}\label{tab:LECs}
\begin{tabular}{l c | r c}
\hline\hline
 & LEC & Value & Source \\
 \hline
$\mathcal{L}_{ N}^{(2)}$& $\widetilde{c}_6$ & $5.07\pm 0.15$ & $\mu_p$ and $\mu_n$~\cite{Bauer:2012pv,Yao:2018pzc,Tanabashi:2018oca} \\
 & $\widetilde{c}_7$ & $-2.68 \pm 0.08$ & $\mu_p$ and $\mu_n$~\cite{Bauer:2012pv,Yao:2019avf,Tanabashi:2018oca} \\
\hline
$\mathcal{L}_{N}^{(3)}$ & $d_6$ & $-0.70 \; \text{GeV}^{-2}$ & $N$ EM Form factor \cite{Fuchs:2003ir}\\
& $d_7$ & $-0.49 \; \text{GeV}^{-2}$ & $N$ EM Form factor\cite{Fuchs:2003ir}\\
& $d_{18}$ & $-0.02 \pm 0.08 \; \text{GeV}^{-2}$ & $\pi N$~scattering~\cite{Alarcon:2012kn} \\
\hline
$\mathcal{L}_{\pi\pi}^{(4)}$ & $l_6$ & $(-1.34 \pm 0.12) \times 10^{-2}$ & $\langle r^2 \rangle_\pi$ \cite{Yao:2018pzc}  \\
\hline
$\mathcal{L}_{\Delta N \pi}^{(1)}$ & $h_A$ & $2.87\pm 0.03$ &$\Gamma_\Delta^{\rm strong}$~\cite{Bernard:2012hb} \\
\hline
$\mathcal{L}_{\Delta N \gamma}^{(2)}$ & $g_M$ & $3.16\pm0.16$ & $\Gamma_\Delta^{\rm EM}$~\cite{Blin:2015era} \\
\hline\hline
\end{tabular}
\end{table}
Many of the LECs appearing in the Lagrangian have been obtained from the study of other processes or physical quantities\footnote{Note that the LECs in Table~\ref{tab:LECs} were obtained within the same framework used here, in a full $\mathcal{O}(p^3)$ calculation in the EOMS scheme and, when appropriate, with explicit $\Delta$ using the $\delta$-counting. }. In the lowest order Lagrangian, $\mathcal{L}_N^{(1)}$, the chiral quantities $\widetilde{g}$, $F$, $\widetilde{m}$ and $M$ are expressed in terms of their corresponding physical values, see Appendix, 
Sec.~\ref{app:chiralexpansions}. For the leading order Lagrangian and the rest of physical quantities we take
$F_\pi=92.42$ MeV, $g_A=1.27$, $m_\Delta=1232$ MeV and $e^2=4 \pi /137$. 

In this work, we compare our model with the experimental database and minimize the $\chi^2$, taking as fitting parameters the remaining free LECs.  In particular,  the combination $\{ d_8+d_9 \}$ that appears exclusively in the  $\pi^0 p$ channel, and  the set $\{d_9,d_{20},d_{21},d_{22}\}$ contributing  to the charged pion channels, as shown in Appendix, Sec.~\ref{app:amplitudes}. In Ref.~\cite{Navarro:2019iqj},  $d_{22}$, related to the nucleon axial radius, was fixed from a fit to lattice data at unphysical pion masses~\cite{Yao:2017fym}. However, the quoted error bars might be underestimated~\footnote{See Fig. 4 of Ref.~\cite{Yao:2017fym}, to fully appreciate the uncertainties of that fit.} and we prefer to fix it independently.  Furthermore, in the previous studies of pion photoproduction, its value could not be well assessed because, at $Q^2=0$, its contribution  is fully correlated to that of $d_{21}$. Thus, the inclusion of electroproduction in the current analysis could lead to a more reliable determination of this parameter.

\subsection{Estimation of the observable uncertainties}

We consider two error sources in our calculation of the observables. One comes from the statistical error in the LECs due to the error bars in the experimental data. We propagate the error bars in the fitting LECs to an associated error, $\delta \mathcal{O}_{\text{LECs}}$, for any observable $\mathcal{O}$ through the  relation,
\begin{align}
\delta \mathcal{O}_{\text{LECs}} = & \left( \sum_{i,j} \left[ \text{Corr}(x_i,x_j)\right] \frac{\partial \mathcal{O}(\bar{x}_i)}{\partial x_i} \delta x_i \frac{\partial \mathcal{O}(\bar{x}_j)}{\partial x_j} \delta x_j  \right)^{1/2} ,
\end{align}
where Corr$(x_i,x_j)$ indicates the $(i,j)$-th element of the correlation matrix, giving the estimated correlation among the $x_i$ and $x_j$ LECs. Moreover $\bar{x}_i$, $\delta{x}_i$ refers to the mean and the error values obtained from the fit for any LEC $x_i$.

In addition, another source of error is the systematical error of the theory due to the truncation of the chiral series expansion at a given $\mathcal{O}(p^n)$. We use the method of  Refs.~\cite{Epelbaum:2014efa,Siemens:2016hdi}, namely, for an order $n$  calculation, $\mathcal{O}_{\text{Th}}^{(n)}$, we estimate this systematical error as  
\begin{align}
\delta \mathcal{O}_{\text{Th}}^{(n)} = & \text{max} \left( \left|  \mathcal{O}^{(n_{LO})} \right| B^{n-n_{LO}+1}, \left\{\left| \mathcal{O}^{(k)} - \mathcal{O}^{(l)} \right| B^{n-l} \right\} \right), \hspace{1cm} n_{LO} \leq l \leq k \leq n.
\label{eq:syserror}
\end{align}
We take $B=m_\pi/\Lambda_b$ and $\Lambda_b$ the breakdown scale of the chiral expansion, $\Lambda_b= 4\pi F_\pi \sim 1$ GeV as in Ref.~\cite{Yao:2017fym}. In the present work we have $n_{LO}=1$ as the lowest order and the upper order is $n=3$.

\subsection{Experimental database}

We compare our model to the available experimental data with some kinematical limits to ensure small external momenta  while staying well below the $\Delta(1232)$ resonance peak. Thus, we have taken the invariant energy of the $\pi N$ system ranging from  threshold  up to $1130$ MeV. Furthermore, from the study of the nucleon electromagnetic form factors ~\cite{Kubis:2000zd,Bauer:2012pv} it is known that a good description beyond $Q^2\sim 0.2$ GeV$^2$ requires the inclusion of vector mesons in the model. Therefore, we have  selected data with transfer momentum, $Q^2<0.15$ GeV$^2$. 
In particular, the case for $Q^2=0$ corresponds to  pion photoproduction.
We expect the  $\mathcal{O}(p^3)$ ChPT calculation with explicit $\Delta$'s to be well suited for the description of the phenomenology in this kinematical region.

\subsubsection{Electroproduction}

The largest amount of data corresponds to  the $\gamma^* p \rightarrow \pi^0 p$ channel.
 Specifically, from the late nineties, we include data for the virtual angular cross section $d \sigma_v / d \Omega_\pi$ at  $Q^2=0.1 \; \text{GeV}^2$, obtained by the Amsterdam Pulse Stretcher facility~\cite{vandenBrink:1997cs},  and data from MAMI \cite{Distler:1998ae} for the observables $d \sigma_{TT}/d \Omega_\pi$, $d \sigma_{TL}/d \Omega_\pi$ and the combination $\left( d \sigma_{T}/d \Omega_\pi + \varepsilon d \sigma_{L}/d \Omega_\pi \right)$. Later, very precise energy dependence data has been obtained at  $Q^2=0.05\; \text{GeV}^2$ in Mainz~\cite{Weis:2007kf} for the observables $d\sigma_{TT}/d \Omega_\pi$, $d\sigma_{TL}/d \Omega_\pi$, $\left( d \sigma_{T}/d \Omega_\pi + \varepsilon d \sigma_{L}/d \Omega_\pi \right)$ and the asymmetry $A_{TLP'}$.
More recently, data for $d \sigma_{TL}/d \Omega_\pi$ and $\left( d \sigma_{T}/d \Omega_\pi + \varepsilon d \sigma_{L}/d \Omega_\pi \right)$ were published for additional $Q^2$ values~\cite{Merkel:2011cf}.





There are far less data for the pion charged channel $\gamma^*p \rightarrow \pi^+ n$.
Nonetheless, they are crucial to determine  LECs like $d_{20}$ and $d_{21}$. We consider data on  $d\sigma_{T}/d\Omega_\pi$, $\sigma_{L}/d\Omega_\pi$, $d\sigma_{TL}/d\Omega_\pi$ and the total $d \sigma_v/d\Omega_\pi$ at a fixed  $Q^2=0.117$ GeV$^2$ measured at   Mainz~\cite{Blomqvist:1996tx}. Later, the experiment was extended to other $Q^2$  values for $d\sigma_{T}/d\Omega_\pi$, $\sigma_{L}/d\Omega_\pi$ and $d \sigma_v/d\Omega_\pi$~\cite{Liesenfeld:1999mv,Baumann:2005}, and more recently to lower energies~\cite{Friscic:2016tbx}.
%
%
%
%

\subsubsection{Photoproduction}
We extend the database used in Ref.~\cite{Navarro:2019iqj}  with the inclusion of  some recent data.
For the $\gamma p \rightarrow \pi^0 p$ channel, we have added the measurements on transverse polarized protons from Ref.~\cite{Schumann:2015ypa}. They correspond to  the observable $T d \sigma/d\Omega_\pi$ \cite{Schumann:2015ypa}, where $T$ is the target asymmetry and $d \sigma/d\Omega_\pi$ the differential cross section~\cite{Navarro:2019iqj}.
We have also included the  total cross section results for the threshold photoproduction on the neutron from Ref.~\cite{Briscoe:2020qat}.

\section{Results and discussion}

\subsection{Low-Energy-Constants}

The theoretical model has been compared with the full photoproduction and electroproduction database previously introduced, minimizing the $\chi$-squared function by varying the values of the free LECs. In the calculation, we have fixed the LECs   from Table~\ref{tab:LECs} to their central values, except for  $g_M$. We have let the $\gamma \Delta N$ coupling, $g_M$, which proved  of paramount significance in the description of $\pi^0$ photoproduction~\cite{Blin:2014rpa}, to fluctuate around the central value obtained from the electromagnetic $\Delta$ width.

\begin{table*}[t]
\caption{Fit results for the LECs. The coupling $g_M$ is dimensionless and $d_i$ in units of GeV$^{-2}$. }
\label{tab:LECfit}
\begin{tabular}{l |c c c  c c  c | c c c}
 &$d_8 + d_9$ &       $d_8 - d_9$ &   $d_{20}$ &     $d_{21}$ &      $d_{22}$ &    $g_{M}$ &    $\chi^2$  &$\chi^2_\gamma$&$\chi^2_e$  \\
\hline
 Full model&$1.12\pm0.01$& $0.63 \pm0.15$& $-0.29 \pm 0.09$  & $1.64 \pm0.06$& $0.95 \pm0.13$& $2.90 \pm 0.01$& 2.7&1.7&5.1\\
\hline
$\Delta$-less & $3.44\pm0.01$& $4.75 \pm0.18$& $-3.01\pm 0.09$  & $4.50 \pm0.06$& $0.45 \pm0.12$& - & 13.2 &16.8& 4.4\\
\hline
\end{tabular}
\end{table*}

We have chosen to  fit the combinations $\{ d_8 +d_9\}$ and $\{ d_8-d_9\}$, instead of the individual constants, because of the important correlation among $d_8$ and $d_9$. Actually, they  appear  in the amplitudes for  $\pi^0$ production just in the combination $\{ d_8 + d_9\}$, while  the charged $\pi^{\pm}$ channels depend only on $d_9$. Given that the $\pi^0$ processes represent, so far, the most precise and largest  amount of data, the $\{ d_8+d_9\}$ combination can be determined with a higher accuracy. Evidently, better data for  the $\pi^\pm$ channels, would be essential to obtain more precise results for  $d_9$ or, similarly, for $\{ d_8-d_9\}$.

The  parameters $\{ d_{20}, d_{21}, d_{22}\}$ are only relevant for the charged channels $\gamma(^*) p \rightarrow \pi^+ n$ and $\gamma(^*) n \rightarrow \pi^- p$. The relatively low precision of the data and their scarcity limits the precision of their determination. Furthermore, these channels are already rather well described by the lower order predictions and in consequence the $\mathcal{O}(p^3)$ LECs play a  small role.
It is worth mentioning that in photoproduction, $d_{21}$ and $d_{22}$ appear only in the combination $\{2 d_{21}-d_{22}\}$ while for electroproduction that is not anymore the case (see App. Sec.~\ref{app:amplitudes}). Therefore, the full correlation is broken once  electroproduction is considered in the fit.

Clearly,  pion electroproduction reactions probe the $Q^2$  dependence of the scattering amplitude. Thus, it allows for the exploration  of  LECs like $\{ d_6,d_7,l_6\}$, which are relevant for the description of the nucleon EM form factors and the pion charge radius and which appear in the electroproduction case.

The LECs values obtained by the fit are presented in Table~\ref{tab:LECfit},  together with the full $\chi^2$ per degree of freedom and the partial contributions of photo-($\chi^2_\gamma$) and electroproduction ($\chi^2_e$).  All the fitted $d_i$'s are of natural size and, thus, the contribution of the associated mechanisms is relatively small at low energies.  While the global result is acceptable, as it will be better shown in the detailed comparison with various observables, it is clear that the model reproduces to a greater degree the photoproduction data. 

The results for $g_M$ and  $\{ d_8 +d_9\}$ agree well with those obtained in the analysis of Ref.~\cite{Navarro:2019iqj}, which studied photoproduction within the same framework but imposed full isospin symmetry on the loop calculation. Our change, using physical masses in the loops, has led to a substantially lower $\chi^2_\gamma$ value and to some small changes in $\{ d_8 -d_9\}$ and  $d_{20}$. A larger variation can be observed in $d_{21}$ and  $d_{22}$ but this could be deceptive. The photoproduction amplitude only depends on the combination $\{2d_{21}-d_{22}\}$, which it has changed little. The separation of the two constants made in Ref.~\cite{Navarro:2019iqj} was based on the use of   $d_{22}=5.20 \; \text{GeV}^{-2}$, taken from Ref.~\cite{Yao:2017fym}. This value, obtained from lattice and already discussed, is clearly disfavoured by the electroproduction data. However, our result is close to an alternative fit of Ref.~\cite{Yao:2017fym} that restricted lattice data to low $Q^2$ values.

All the fitted $d_i$'s appear in the evaluation of neutrino induced pion production off nucleons and could be used to improve the corresponding predictions. This is specially  important in the current precision era of neutrino physics, where an adequate modelling of cross sections and backgrounds is necessary for the investigation of neutrino masses, mixing angles and other properties~\cite{Alvarez-Ruso:2014bla}.
Our  results give support to the first ChPT calculations of these weak production processes~\cite{Yao:2018pzc,Yao:2019avf}, which assumed a natural size for these parameters to estimate the uncertainties of the theoretical predictions.

\subsection{Electroproduction observables}

\begin{figure}    
    \includegraphics[page=1,angle=0,scale=0.5]{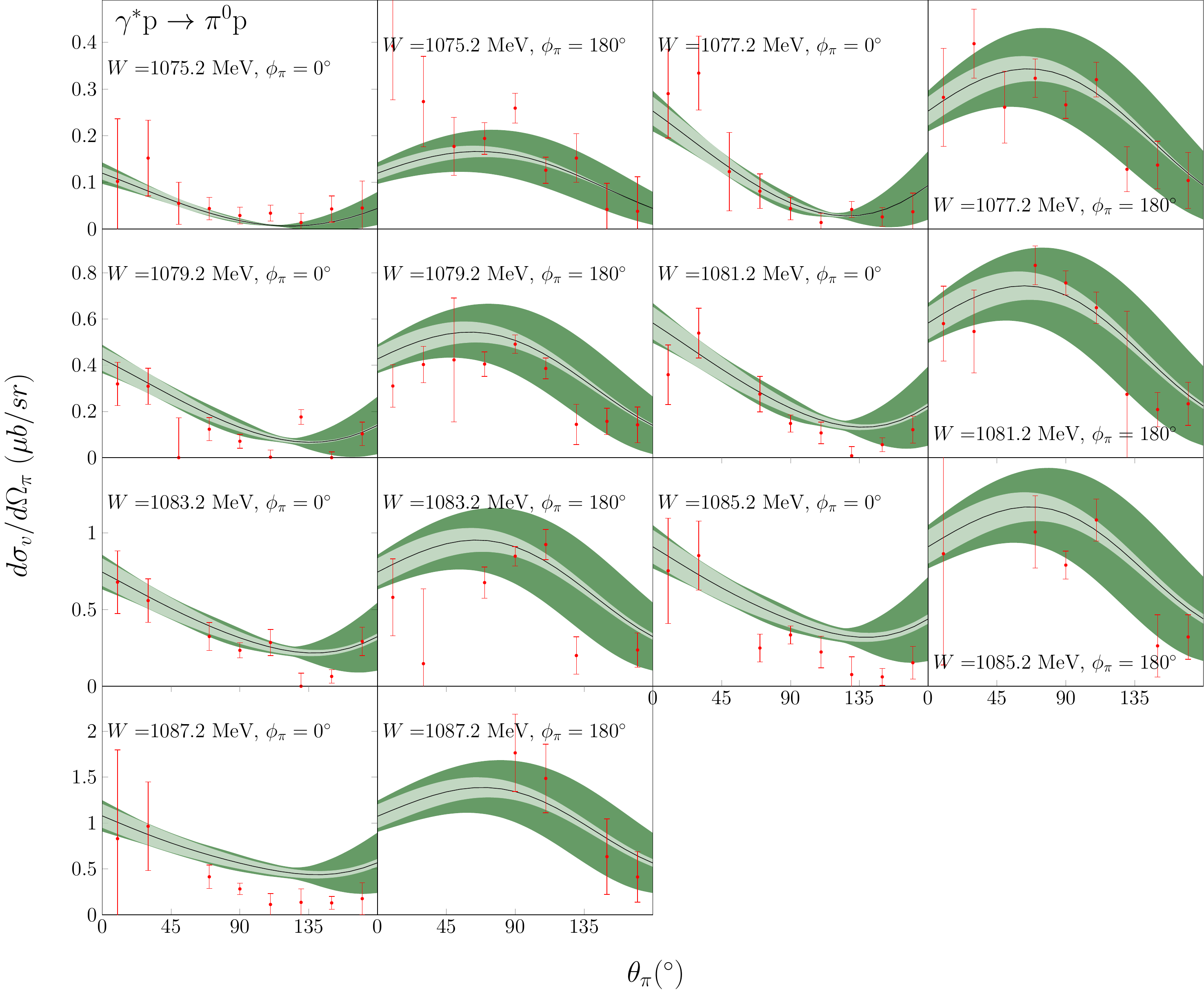}
\caption{Angular distribution of the virtual cross section $d\sigma_v/d\Omega_\pi$ at different angles and energies, transfer momentum $Q^2=0.10\;\text{GeV}^2$ and virtual-photon polarization $\varepsilon=0.670$. Solid line shows the theoretical results, the inner band depicts the statistical error from the LECs variation within 1-$\sigma$ as in Table~\ref{tab:LECfit}. The outer band represents the total error including the systematical error from chiral truncation, Eq.~(\ref{eq:syserror}), added to the statistical one in quadrature. Data from Ref.~\cite{vandenBrink:1997cs}.}    
\label{fig:electroDSGpi0p}
\end{figure}

\begin{figure}
    \includegraphics[page=1,angle=0,scale=0.5]{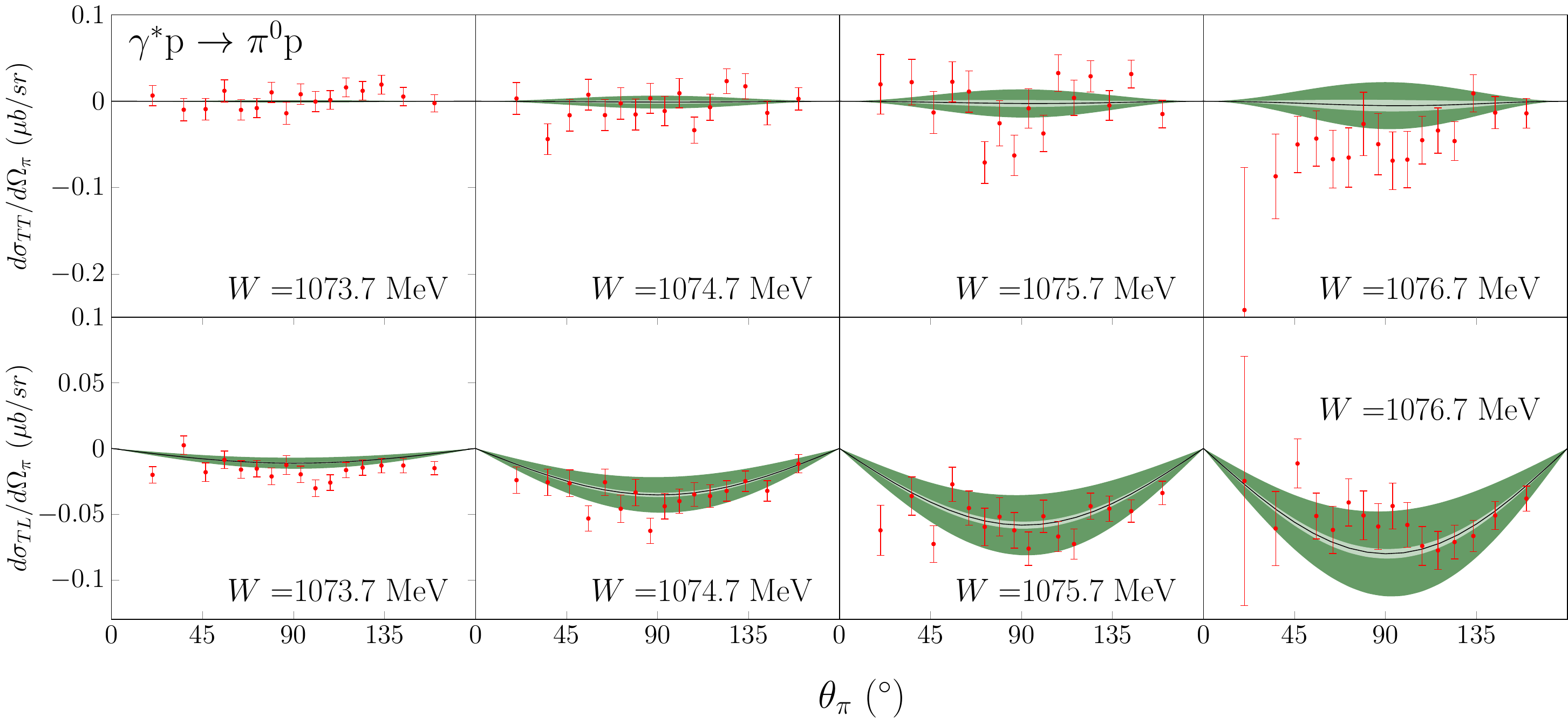}
\caption{ $d\sigma_{TT}/d\Omega_\pi$ and  $d\sigma_{TL}/d\Omega_\pi$ as function of the c.m. pion angle $\theta_{\pi}$ for the channel $\gamma^* p \rightarrow \pi^0 p$ at $Q^2=0.1\;\text{GeV}^2$ and with polarization $\varepsilon=0.713$. Data  from Ref.~\cite{Distler:1998ae}. Description as in Fig.~\ref{fig:electroDSGpi0p}.}    
\label{fig:electroDTTpi0p}
\end{figure}

\subsubsection{$\gamma^* p \rightarrow \pi^0 p$ channel}
\begin{figure}    
    \includegraphics[page=1,angle=0,scale=0.62]{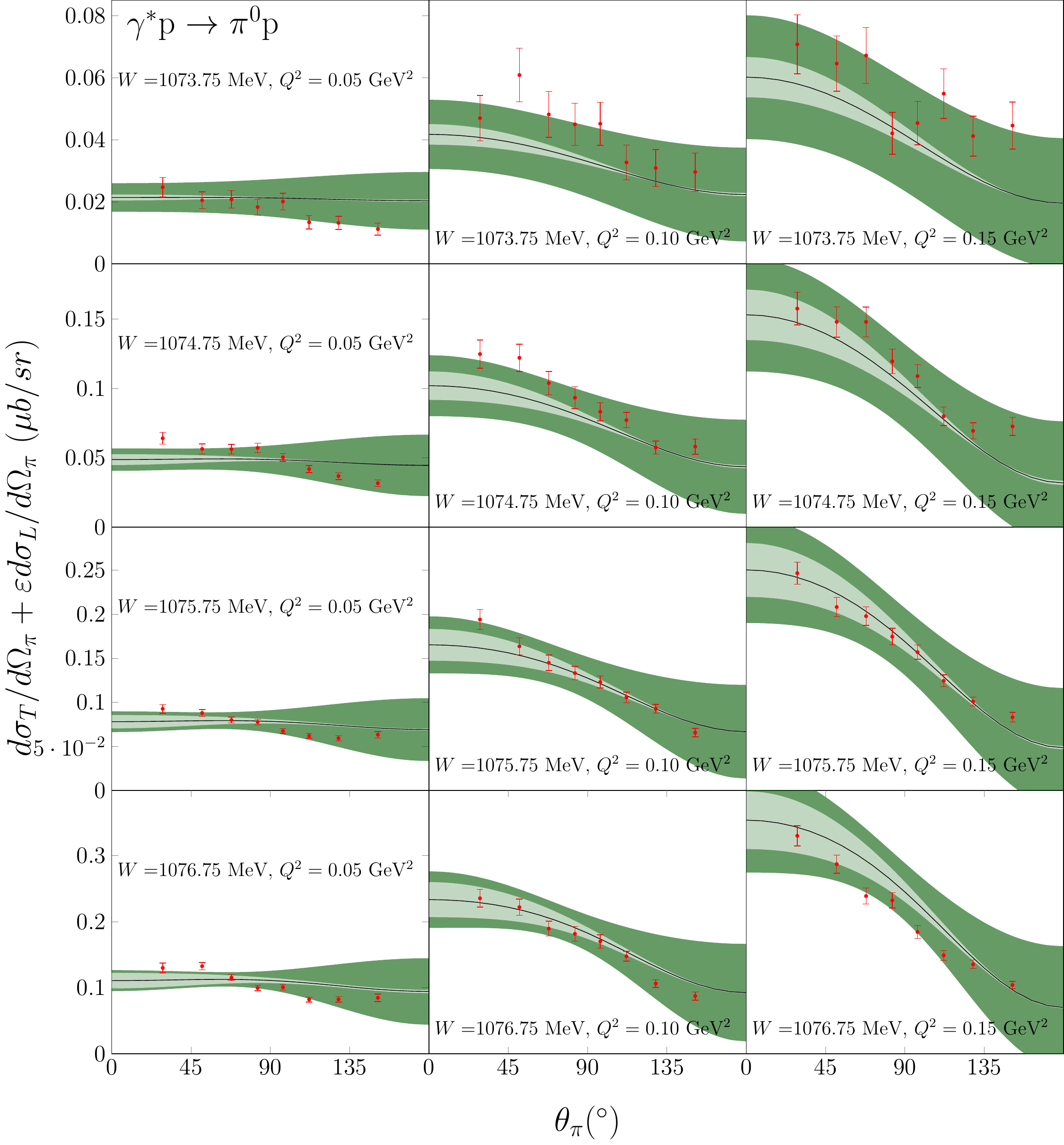}
\caption{Angular distribution for $d\sigma_T + \varepsilon d\sigma_L$ at different c.m. energy values, $W$. The transfer momenta at $Q^2=0.05\;\text{GeV}^2$ corresponds to polarization values of $\varepsilon=0.932$, $Q^2=0.10\;\text{GeV}^2$ to $\varepsilon=0.882$ and $Q^2=0.15\;\text{GeV}^2$ to $ \varepsilon=0.829$. Data from \cite{Merkel:2011cf} and description as in Fig.\ref{fig:electroDSGpi0p}.}    
\label{fig:electroDSTLpi0p2}
\end{figure}

\begin{figure}    
    \includegraphics[page=1,angle=0,scale=0.5]{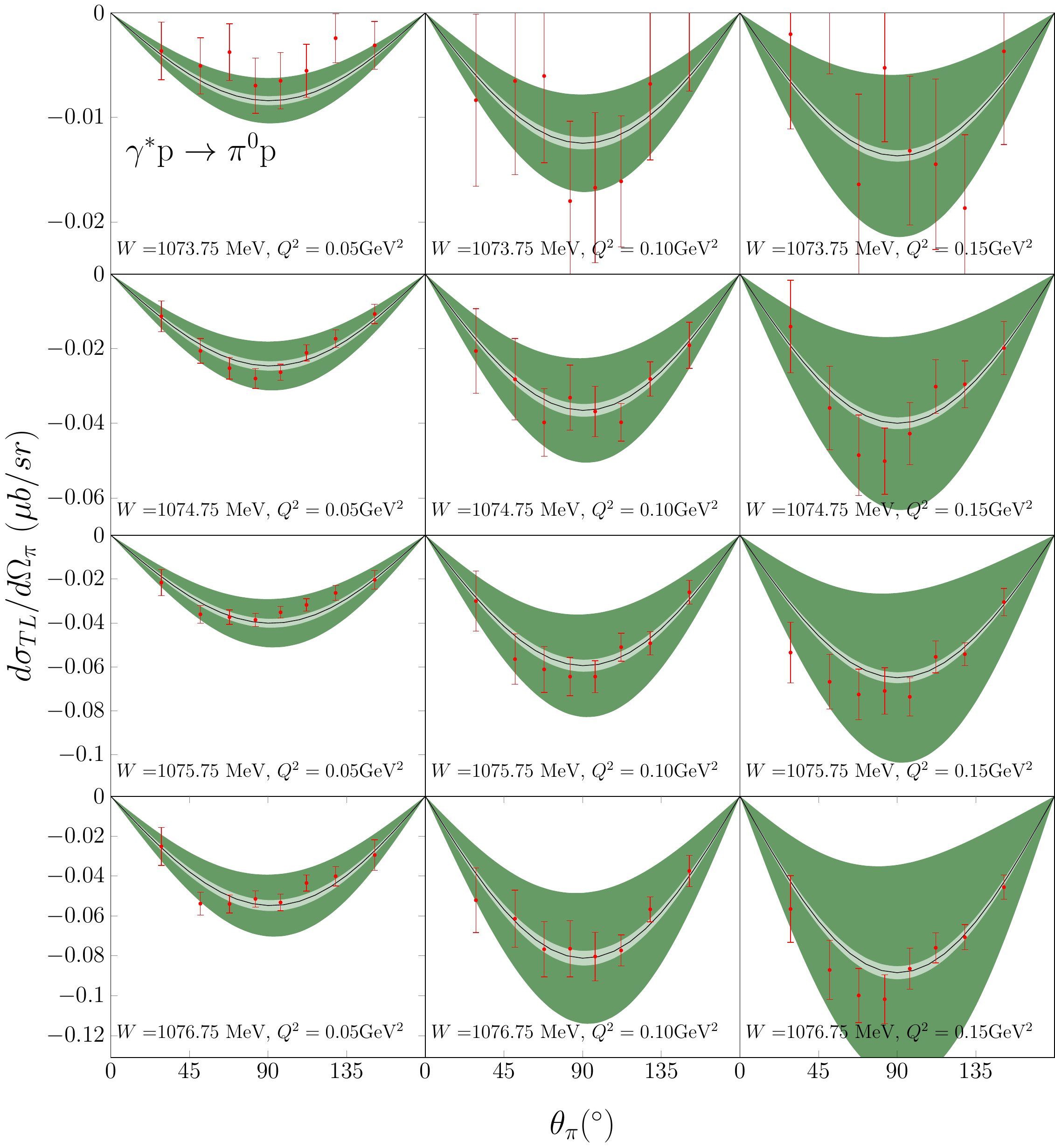}
\caption{Angular distribution for $d\sigma_{TL}/d\Omega_\pi$ for  different c.m. energy values, W. The transfer momenta at $Q^2=0.05\;\text{GeV}^2$ corresponds to polarization values of $\varepsilon=0.932$, $Q^2=0.10\;\text{GeV}^2$ to $\varepsilon=0.882$ and $Q^2=0.15\;\text{GeV}^2$ to $ \varepsilon=0.829$. Data from \cite{Merkel:2011cf} and description as in Fig.~\ref{fig:electroDSGpi0p}.}    
\label{fig:electroDTLpi0p2}
\end{figure}

In this section, we show our results for the $\pi$ electroproduction process compared to the experimental data. We start with the $\gamma^* p \rightarrow \pi^0 p$ channel, that represents the largest amount  of data, in Figs.~\ref{fig:electroDSGpi0p}-\ref{fig:electroDTLpi0p2}. We should remark that, among the third order fitted LECs, this channel's amplitude depends only on the $\{d_8+d_9\}$ combination, that is much constrained by   neutral pion photoproduction.  Actually, the current fit results for that LEC are fully consistent with the previous determination based just on photoproduction~\cite{Navarro:2019iqj}. Overall, the agreement with data is good  for all the  observables considered here.

In Fig.~\ref{fig:electroDSGpi0p}, we show the virtual photon cross section, $d\sigma_v/d\Omega_\pi$, at  several energy bins close to threshold,  $Q^2=0.10\; \text{GeV}^2$ and for $\varepsilon=0.67$, compared to the NIKHEF data from Ref.~\cite{vandenBrink:1997cs}. The angular dependence,  on both $\theta_{\pi}$ and $\phi_{\pi}$, and the energy dependence are well reproduced. 

The various pieces,  related to the longitudinal and transverse responses and their interference, which contribute to the total cross section of Eq.~(\ref{eq:virtualsigma}),  are explored next.   In Fig.~\ref{fig:electroDTTpi0p}, we compare the model with the angular dependence of  $\sigma_{TT}$ and $\sigma_{TL}$ measured by MAMI~\cite{Distler:1998ae} at several energies very close to threshold. The two observables are very small. Both the size and the energy dependence are well accounted for by our calculation.  Much larger is the observable $d\sigma_T/d\Omega_\pi + \varepsilon d\sigma_L/d\Omega_\pi$ from a much more recent MAMI experiment~\cite{Merkel:2011cf} and depicted in Fig.~\ref{fig:electroDSTLpi0p2}.  These latter results show the $Q^2$ dependence, that at the low energies involved and for the relatively small $Q^2$ values is well described by the model.

The $Q^2$ dependence is also explored for $d\sigma_{TL}$ in Fig.~\ref{fig:electroDTLpi0p2}, which also shows a good agreement for the angular distribution at several $Q^2$ values. We should remark that for neutral pions, apart from the fixed LECs, this dependence is only sensitive to $\{d_8+d_9\}$ and $g_M$, which are strongly constrained by the photoproduction ($Q^2=0$) data.

\begin{figure}    
    \includegraphics[page=1,angle=0,scale=0.7]{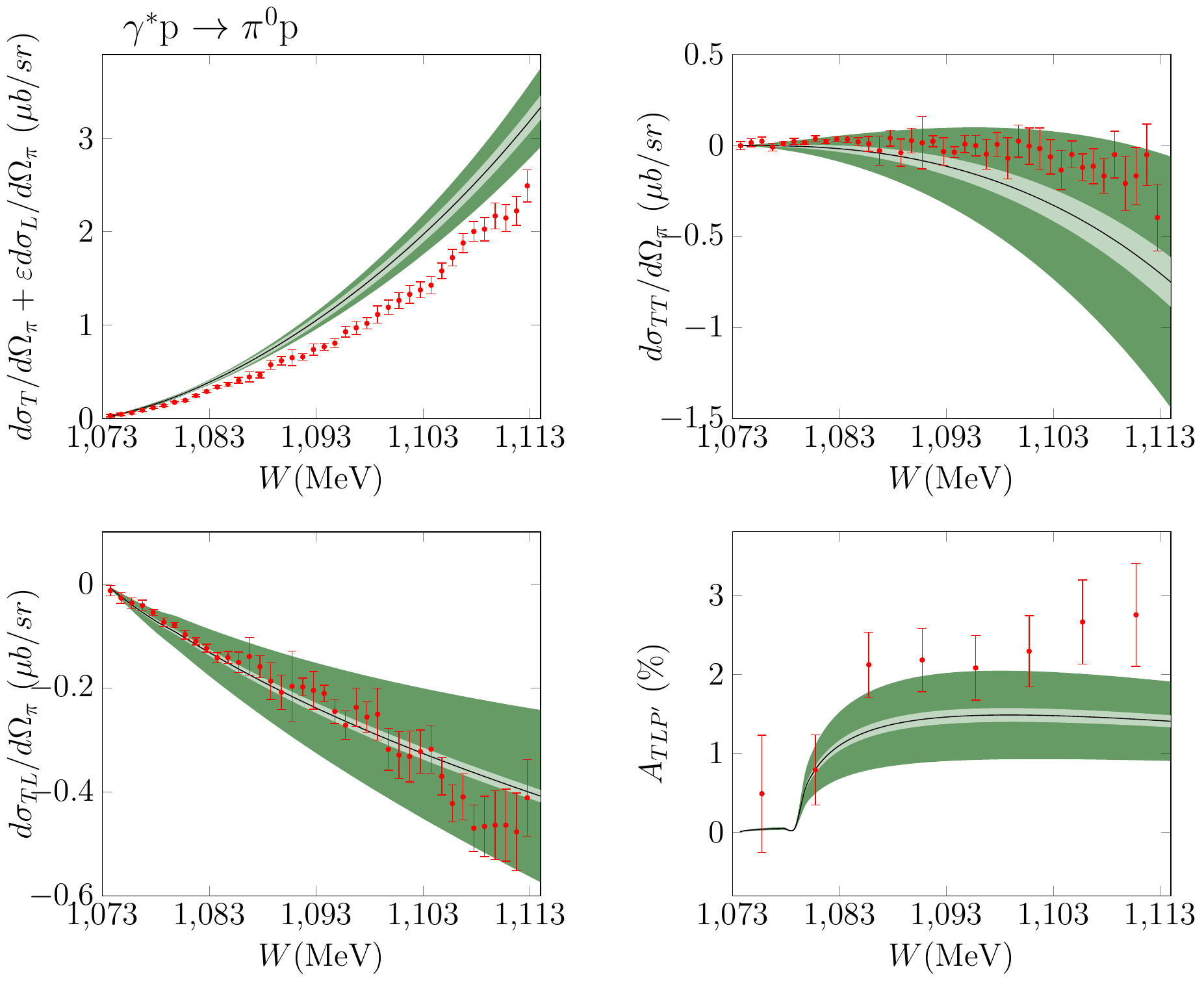}
\caption{Energy dependence for $d\sigma_T + \varepsilon d\sigma_L$, $d\sigma_{TT}$, $d\sigma_{TL}$ and $A_{LT'}$ at $Q^2=0.05 \;\text{GeV}^2$, $\varepsilon=0.933$, $\theta_\pi=90^\circ$. Data from \cite{Weis:2007kf}.}    
\label{fig:electroATLPpi0p2}
\end{figure}

Finally, in Fig.~\ref{fig:electroATLPpi0p2},  we compare our calculation with the very copious and precise data of Ref.~\cite{Weis:2007kf}, where the energy dependence of 
 $d\sigma_T$, $d\sigma_{TT}$, $d\sigma_{TL}$ and $A_{LT'}$ has been investigated at  $Q^2=0.05\;\text{GeV}^2$ and photon transverse polarization $\varepsilon=0.933$.    
For $d\sigma_T/d\Omega_\pi +\varepsilon d\sigma_L/d\Omega_\pi$ and $d\sigma_{TT}/d\Omega_\pi$, the calculation agrees well up to a few MeV above threshold, what is consistent with the results shown in Fig.~\ref{fig:electroDSTLpi0p2}. However, we overestimate the absolute value of the observable at higher energies. In fact, our fit curve behaves as  the HBChPT result of Ref.~\cite{Bernard:1996bi} discussed in~\cite{Weis:2007kf}. The agreement with $\sigma_{TT}$  is good and with $\sigma_{TL}$ excellent, in both cases improving the HBChPT prediction.  In these three cases, the quality of the agreement of our $\mathcal{O}(p^3)$ model is very similar to that of the  $\mathcal{O}(p^4)$ $\Delta$-less  covariant ChPT calculation of Ref.~\cite{Hilt:2013fda}.

 Also well reproduced is the beam helicity asymmetry,  $A_{LT'}$, a quite small effect, which shows the cusp related to the $n\pi^+$ threshold.  The use of the physical masses in the loops, and the corresponding isospin symmetry breaking is essential for a proper reproduction of this shape.

Summarizing, the theoretical results  for the $\pi^0$ channel are in accordance with data, describing properly the angular dependence and  the $Q^2$ evolution.  In regard to the energy, we obtain the best results  very close to threshold.  Nonetheless, the model starts to overestimate data for the observable $d\sigma_T + \varepsilon d\sigma_L$ at  higher energies, see Fig.~\ref{fig:electroATLPpi0p2}. Actually, this observable contributes strongly to the total $\chi^2$. On the other hand, it is very sensitive to  $c_6+c_7$, $2d_7+d_6$ and $g_M$, which were restricted to the values allowed by the study of other processes. In our calculation, the only totally free parameter relevant for this channel has been  the combination $\{d_8+d_9\}$, strongly constrained by the abundant photoproduction data.

\subsubsection{$\gamma^* p \rightarrow \pi^+ n$ channel}

The channel $\gamma^* p \rightarrow \pi^+ n$ depends on the $\mathcal{O}(p^3)$ LECs  $d_9$, $d_{20}$, $d_{21}$ and $d_{22}$, as well as the $\mathcal{O}(p^4)$ one $l_6$.\footnote{Other $\mathcal{O}(p^4)$ LECs appearing in the tree-level amplitudes for the $\gamma^* p \rightarrow \pi^+ n$ channel are $l_3$ and $l_4$. However, they are cancelled in the amplitude expansion up to $\mathcal{O}(p^3)$ when, at the same time, we introduce the pion wave function renormalization, $\mathcal{Z}_\pi$, and the pion chiral mass, $M$, as a function of the pion physical mass, $M_\pi$. See App. Sec.~\ref{app:wfr} and Sec.~\ref{app:chiralexpansions}.}  Thus, there are more fitting LECs than for the neutral pion channel. Furthermore, the data are scarce. For these reasons, there are less constrains on the relevant LECs and the statistical error is considerably wider.

\begin{figure}    
    \includegraphics[page=1,angle=0,scale=0.5]{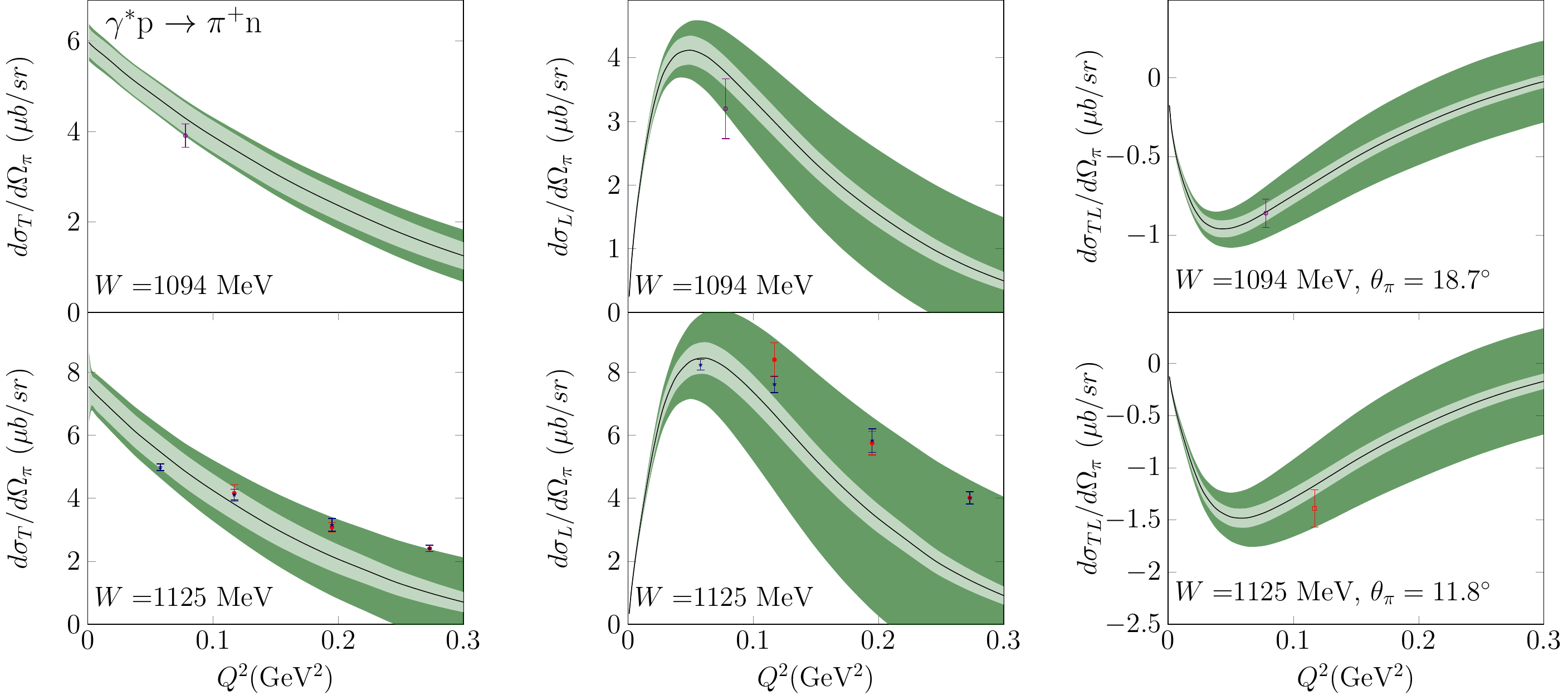}
\caption{$d\sigma_T$, $d\sigma_L$ and $d\sigma_{TL}$ as functions of $Q^2$ for the $\gamma^* p \rightarrow \pi^+ n$ process. For  $d\sigma_T$ and $d\sigma_L$, the pion angle is $\theta_\pi=0^\circ$. Magenta circles: data from \cite{Friscic:2016tbx}. Red squares: data from \cite{Blomqvist:1996tx,Liesenfeld:1999mv}. Blue dots: data from \cite{Baumann:2005}.}  
\label{fig:electroDTpipn}
\end{figure}

We find that the few and scattered virtual photon cross section data~\cite{Blomqvist:1996tx,Liesenfeld:1999mv} agree well, within errors, with the theoretical model, and that the $\pi^+$ channel is more sensitive to the lower orders than to the $\mathcal{O}(p^3)$ contributions. 
In Fig.~\ref{fig:electroDTpipn}, we present  $d\sigma_T$, $d\sigma_L$ and $d\sigma_{TL}$ as a function of $Q^2$ at various pion angles and from several experiments that are also well reproduced.

\subsection{Photoproduction}

The use of physical masses in the loop propagators and, therefore, the breaking of the isospin symmetry is the main difference of this calculation with Refs.~\cite{Blin:2016itn,Navarro:2019iqj}. It leads to a better description of the low energy region, where the effects of the different masses and thresholds are more relevant. Furthermore, in Refs.~\cite{Blin:2016itn,Navarro:2019iqj}, there was a systematic overestimation of the cross section at backward angles for the $\pi^0 p$ channel at all energies. The breaking of the isospin symmetry in the loops has now much improved the agreement with that cross section. As a consequence, the partial $\chi^2$, considering only photoproduction, has been reduced from 3.2 to 1.5.  Also, without isospin breaking, the fit prefers values  of 
$d_{18}$ large and positive, which are inconsistent with $\pi N$ scattering. Now, the tension is much reduced and the $\chi^2$ depends less strongly on that parameter.
In the following, we present our results putting emphasis on the comparison with the new data, added to the database after Ref.~\cite{Navarro:2019iqj}, and in the low energy region, that had not  been included in the previous fit.

The $\gamma p \rightarrow \pi^0 p$ channel is the most richly represented in the database, both in the amount and  the precision of  data.
Thus, the relevant LECs, in particular the  $d_8+d_9$ combination, are strongly constrained and get a  relatively small uncertainty in the fit.
\begin{figure}
\includegraphics[page=1,angle=0,scale=0.42]{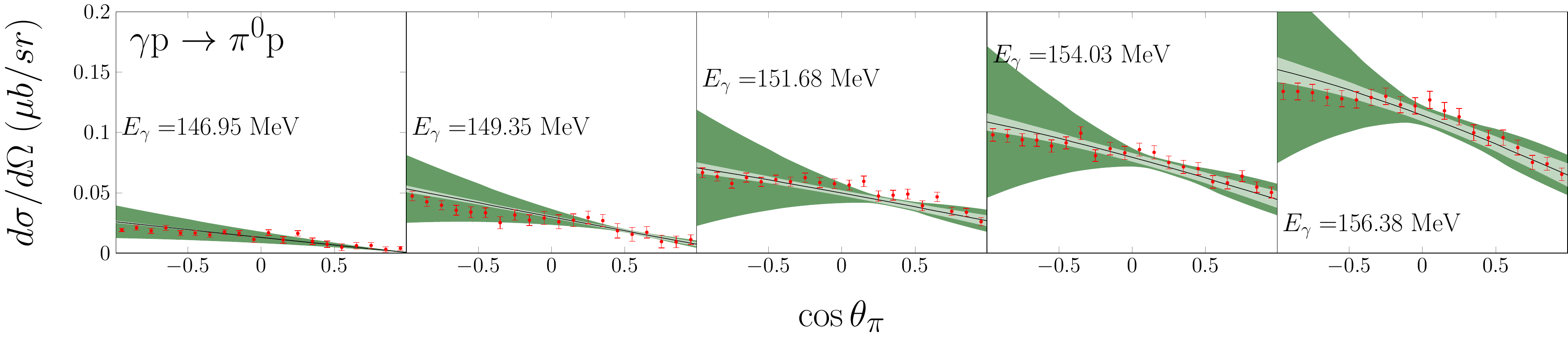}
\caption{Angular cross section for the channel $\gamma p \rightarrow \pi^0 p$. Data from  Ref.~\cite{Hornidge:2012ca}.}
\label{fig:photoDSGpi0p}
\end{figure}
In Fig.~\ref{fig:photoDSGpi0p}, we show the near threshold region for the angular distribution and in Fig.~\ref{fig:photoSGTpi0p} the integrated total cross section $\sigma$ as function of the energy. 
\begin{figure}
\includegraphics[page=1,angle=0,scale=1.0]{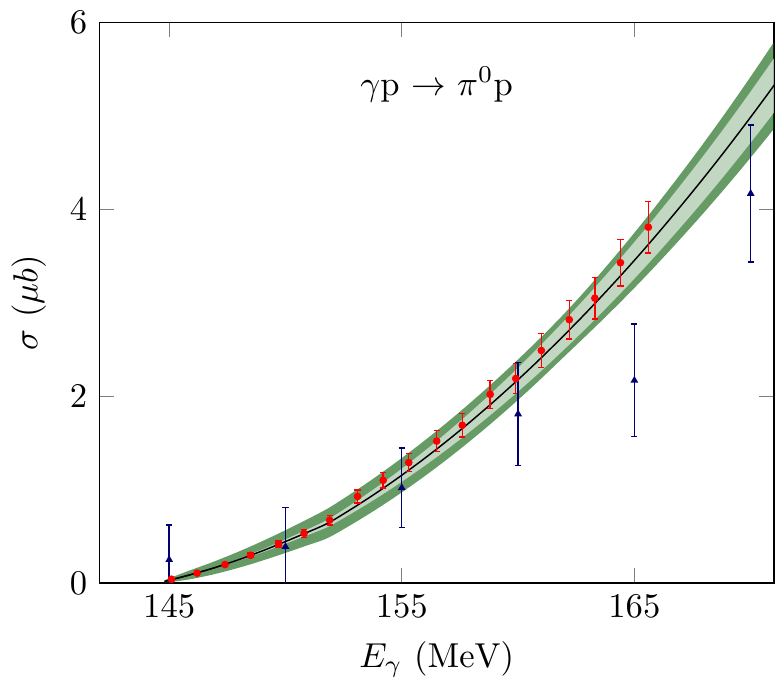}
\caption{Cross section close to threshold for $\gamma p \rightarrow \pi^0 p$. Red circles: data from Ref.~\cite{Schmidt:2001vg}, blue triangles: data from  Ref.~\cite{Schumann:2010js}, not included in the fit.}
\label{fig:photoSGTpi0p}
\end{figure}
Both are well reproduced.
Our calculation still preserves the excellent results for the energy dependence of the total cross section  and for the beam asymmetry  as in the previous work~\cite{Navarro:2019iqj}.
\begin{figure}
\centering
\includegraphics[page=1,angle=0,scale=0.35]{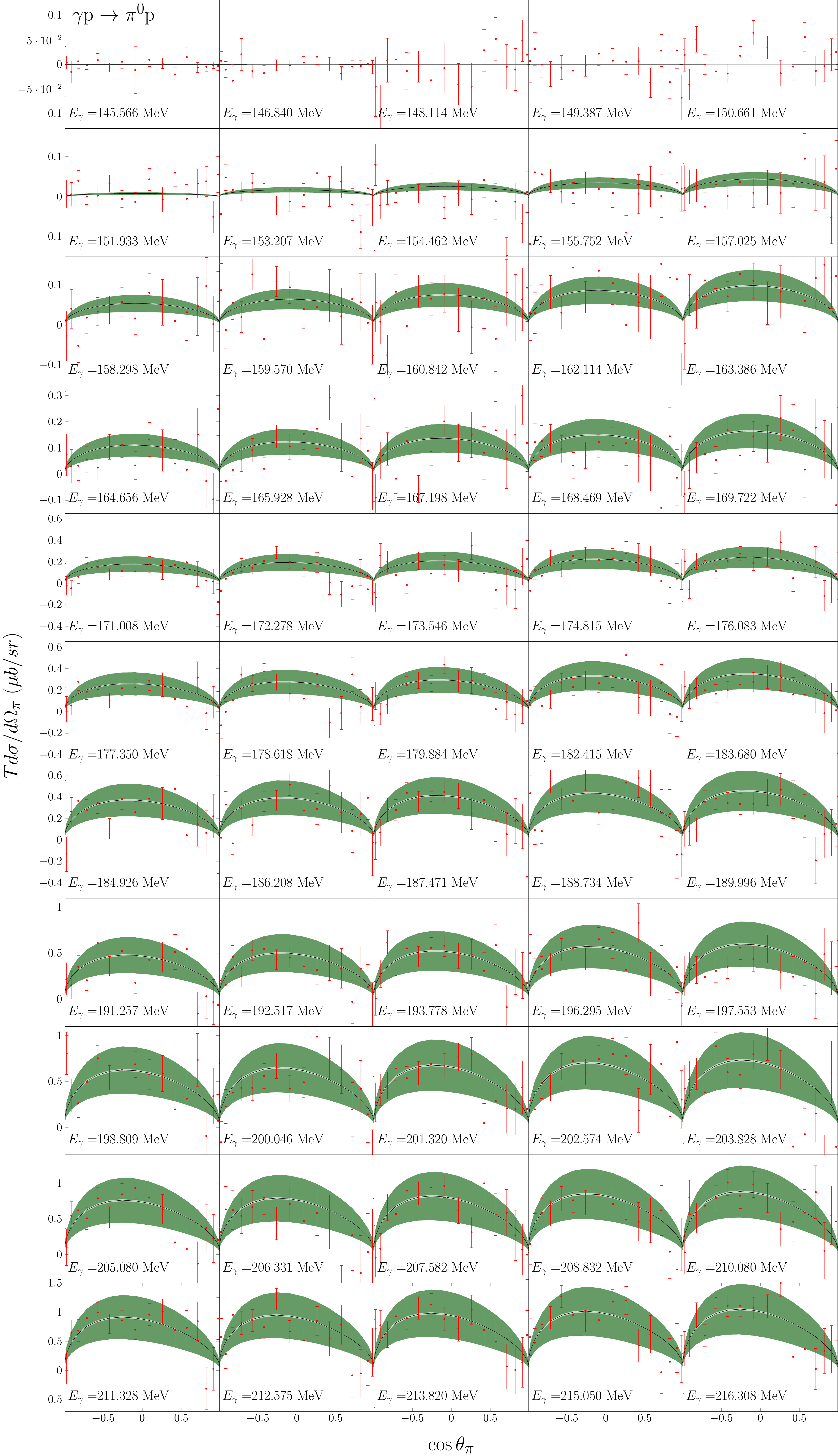}
\caption{Angular distribution of  $T d\sigma/d\Omega_\pi$ for the $\gamma p \rightarrow \pi^0 p$ channel. Data from \cite{Schumann:2015ypa}.}
\label{fig:photoTpi0p}
\end{figure}
In addition, for the $\pi^0p$ channel, we have analyzed the data from Ref.~\cite{Schumann:2015ypa} studying the process occurring on transversely polarized protons.  The observable $T d\sigma/d\Omega_\pi$ is sensitive to the cusp effects due to the $n\pi^+$ threshold.
The results are shown in Fig.~\ref{fig:photoTpi0p}, with $T$ and present a good agreement for the full range of energies.

\begin{figure}
    \centering
\includegraphics[page=1,angle=0,scale=1.0]{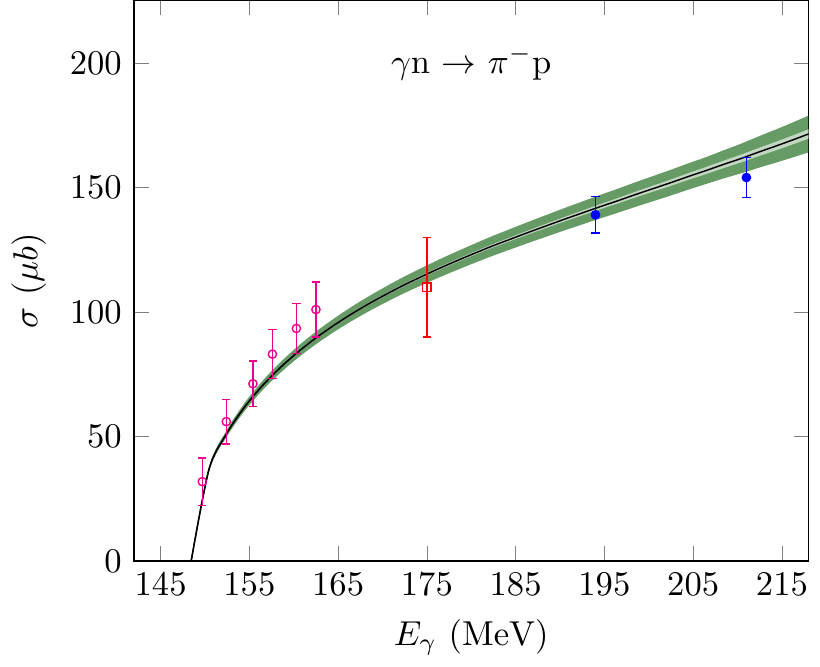}
    \caption{Cross section for the $\gamma n \rightarrow \pi^- p$ process. Data from \cite{Briscoe:2020qat} in magenta circles; red squares, data from \cite{White:1960ukk} and blue dots, data from \cite{WA92} (not included in the fit). }
    \label{fig:13foto}
\end{figure}

The quality of the  agreement with the channels with charged pions has also improved upon Ref.~\cite{Navarro:2019iqj}, as can be seen comparing the partial $\chi^2$'s. We would like to emphasize the recent results, shown in Fig.~\ref{fig:13foto}, for the $\gamma n \rightarrow \pi^- p$ process~\cite{Briscoe:2020qat} very close to threshold. They  have considerably enriched the database for this channel and therefore lead to a better determination of the LECs relevant for this channel, $d_9$, $d_{20}$ and the combination  $2d_{21}-d_{22}$.

\subsection{$\Delta$ contribution}
To explore the importance of the inclusion of the explicit $\Delta(1232)$ in the model, we repeated the fit without the corresponding mechanisms. The results for the LECs and $\chi^2$ are shown in the second row of Table~\ref{tab:LECfit}. It is remarkable that the $\Delta$ contribution, which depends only on well constrained parameters, ($h_A$ and  $g_M$), improves substantially the global agreement with data. It is also noteworthy that most of the fitted $d_i$ LECs are much larger in the $\Delta$-less case, indicating the need of a more important third order and a slower chiral convergence.
Comparing with the full model, we see that, with the current data set, the $\chi^2$ for photoproduction is considerably worsened, whereas for electroproduction $\chi^2$ is little modified, even showing a little improvement. 
In particular, we have found that $\Delta$ inclusion worsens the overestimation for $d\sigma_{T}/d\Omega_\pi$ in Fig.~\ref{fig:electroATLPpi0p2}. However, it improves the  agreement with the other observables of the same figure. This point is relevant, because that observable has the largest, may be excessive, weight in the $\chi^2$ calculation among the full  electroproduction data set, followed by $d\sigma_{TT}/d\Omega_\pi$ from the same experiment \cite{Weis:2007kf}. This is due to the large number of points and their precision.

\begin{figure}
\includegraphics[page=1,angle=0,scale=1.0]{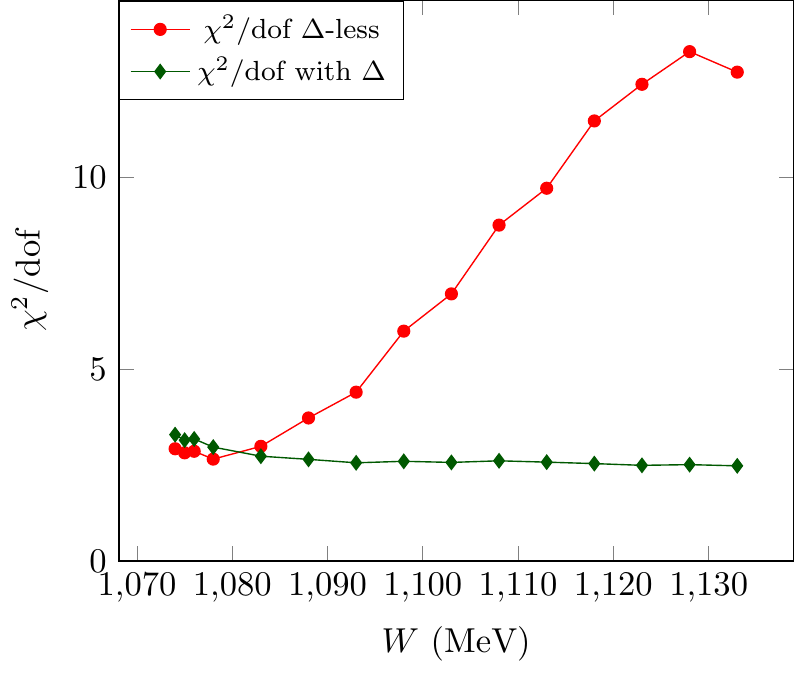}
\caption{$\chi^2/$dof  as function of the maximum $W$ considered in the fitting procedure. Full model at $\mathcal{O}(p^3)$ with $\Delta$ (green diamonds) and without $\Delta$ (red circles).} 
\label{fig:chi2vsW}
\end{figure}

In contrast, the $\Delta$ role in photoproduction is of the uttermost importance to reproduce the energy dependence of data. The $\Delta$-less model is unable to describe the energy evolution of the cross sections, mostly in the $\pi^0$ channel, even with the inclusion of the $\mathcal{O}(p^3)$ one-loop amplitudes.
This failure can be appreciated in Fig.~\ref{fig:chi2vsW}. There, we show the $\chi^2$ per degree of freedom as a function of the maximum invariant energy, $W$ considered in the fitting procedure. The quality of the agreement remains stable for the full model whereas without  explicit  $\Delta$ the $\chi^2$ function grows fast as a function of the energy  and it is impossible to describe data at this chiral order.

\section{Summary}

In this work, we have studied  pion production off the nucleon induced by virtual and real photons at low energies. We have made a full $\mathcal{O}(p^3)$ calculation, in the $\delta$ counting,  in covariant ChPT including explicitly the $\Delta(1232)$ resonance and employing the EOMS renormalization scheme.  
The free LECs of the theoretical model have been fixed by fitting it to the available pion electro- and photoproduction data.   We have considered a restricted kinematical region with $\sqrt{s}<1.13$ GeV and $Q^2<0.15$ GeV$^2$, where we expect our model to be reliable and still well below the $\Delta(1232)$ peak.

We have  confirmed the importance of the loop terms. The imaginary parts of the scattering amplitude and the cusp effects, coming from the opening of the various charge channels, are crucial in the description of some low energy observables. To properly  account  for these effects we have used the physical masses of mesons and baryons in the evaluation of the loops, therefore breaking isospin symmetry.

The model describes well all data for  total cross section, angular distributions and  numerous polarization observables. In particular, the agreement is excellent for photoproduction data. In fact, it is better than for other higher order chiral calculations~\cite{FernandezRamirez:2012nw,Hilt:2013fda} that do not include the $\Delta$ resonance. Without $\Delta$, our model is only able to reproduce data a few MeV above threshold.  Neutral pion photoproduction is the most sensitive channel to this resonance due to the smallness of the lower order contributions.

The comprehensive investigation of all electro- and photoproduction channels, including all the available observables, has allowed to disentangle all the relevant third order LECs involved,  $\{ d_8,d_9,d_{20}, d_{21}, d_{22}\}$.
The values obtained for the fitted LECs are all of natural size, what is satisfactory from the point of view of chiral convergence.  Furthermore, this gives support to the uncertainty estimations of recent chiral calculations of neutrino induced pion production. Our results will allow for more precise predictions of the low energy neutrino nucleon cross sections of relevance to achieve the precision goals of modern neutrino experiments.

\section*{Acknowledgments}
We thank M. Ostrick for providing us the data from Ref.~\cite{Schumann:2015ypa}. We also thank A. N. Hiller Blin and De-Liang Yao for useful comments.
G.H.G.N. wishes to thank the Generalitat Valenciana for support in the program  GRISOLIAP-2017-098.  
This research is supported by MINECO (Spain) and the ERDF (European Commission) Grant No. FIS2017-84038-C2-2-P and by the EU Horizon2020 research and innovation programme, STRONG-2020 project, under grant agreement No. 824093.

\section{Appendix}

\subsection{Amplitude parametrizations}

The electromagnetic matrix element between the hadronic states, $\mathcal{M}^\mu$, can be written in terms of the Ball amplitudes~\cite{Ball:1961zza},
\begin{equation}
\mathcal{M}^{\mu} = \bar{u}(p',s') \left( \sum_{i=1}^8 b_i V_i^{\mu} \right) u(p,s)
\end{equation}
with the Ball vector basis, see e.g.~\cite{Hilt:2013fda},
\begin{equation}
\begin{split}
V_1^{\mu} = \gamma^\mu \gamma^5, \;\;&  V_2^{\mu} = P^\mu \gamma^5, \\
V_3^{\mu} = q^\mu \gamma^5, \; \; & V_4^{\mu} = k^\mu \gamma^5, \\
V_5^{\mu} = \gamma^\mu \slashed k \gamma^5, \; \; & V_6^\mu = P^{\mu} \slashed k \gamma^5, \\
V_7^{\mu} =q^\mu \slashed k \gamma^5, \; \; & V_8^\mu = k^\mu \slashed k \gamma^5,
\end{split}
\label{eq:Vbasis}
\end{equation}
where $P^{\mu} = (p + p')^{\mu}/2$. As the current $J^\mu$, from Eq.~\ref{eq:current}, obeys the continuity equation, we also have $k_\mu \mathcal{M}^\mu = 0$, leading us to the following relations,
\begin{equation}
\begin{split}
b_1 =& - b_6 (k \cdot P) - b_7 (k \cdot q) + b_8 Q^2, \\
b_2 =& \frac{1}{k \cdot P} \left( Q^2(b_4 + b_5) - b_3 (k \cdot q) \right).
\end{split}
\end{equation}
These relations are sufficient to impose the gauge invariance in the scattering amplitude. They also reduce from eight to six the independent  elements of the basis $\{V_i\}$. Another common parametrization, in terms of the covariant basis elements $M_i^\mu$, is~\cite{Drechsel:1992pn}
\begin{equation}
\sum_{i=3}^8 b_i V_i^\mu = \sum_{j=1}^{6} A_j M_j^{\mu}\,,
\end{equation}
where
\begin{equation}
\begin{split}
M_1^\mu = & - \frac{i}{2} \gamma^5 \left(\gamma^{\mu} \slashed k - \slashed k \gamma^\mu \right), \\
M_2^\mu = & 2 i \gamma^5 \left( P^\mu k \cdot \left( q - \frac{1}{2} k \right) - \left( q - \frac{1}{2} k \right)^\mu k \cdot P \right), \\
M_3^\mu = & -i \gamma^5 \left( \gamma^\mu (k \cdot q) - q^\mu \slashed k \right), \\
M_4^\mu = & i m_N \gamma^5 (\gamma^\mu \slashed k -\slashed k \gamma^\mu) - 2 i \gamma^5 (\gamma^\mu k \cdot P -P^\mu \slashed k ), \\
M_5^\mu = & i \gamma^5 \left( k^\mu (k \cdot q) + Q^2 q^\mu \right), \\
M_6^\mu = & -i \gamma^5 \left( k^\mu \slashed k + Q^2 \gamma^\mu \right)\,.
\end{split}
\end{equation}
In the case of photoproduction, $Q^2 =0$, and then $\epsilon_\mu M_j^\mu=0$ for $j=5,6$. 

The relations among the above mentioned parametrizations are given by
\begin{equation}
\begin{split}
A_1 =& i (b_5 + b_6 m_N), \\
A_2 = & - \frac{i \left(-b_3 (k \cdot q) + (b_4+b_5) Q^2 \right)}{(k\cdot P)(2 k\cdot q + Q^2)}, \\
A_3 =& i b_7 ,\\
A_4 =& \frac{i b_6}{2},\\
A_5 =& - \frac{i(b_3+2(b_4+b_5))}{2 k \cdot q + Q^2},\\
A_6 =& -i b_8\,.
\end{split}
\label{eq:Acoeff}
\end{equation}

Using the CGLN basis, as in~\cite{Chew:1957tf,Dennery:1961zz},
we can write
\begin{equation}
\epsilon_\mu \mathcal{M}^\mu =\epsilon_\mu \bar{u}(p_f) \left( \sum_{i=1}^6 A_i M_i^\mu \right) u(p_i) = 4 \pi \dfrac{W}{m_N} \chi_f^\dagger  \mathcal{F} \chi_i\, .
\end{equation}

Then, we find, expressed in the CM frame, the relations between the coefficients of both parametrizations as\footnote{See Ref.~\cite{Pasquini:2007fw} for some help in the derivation of these equations.}
\begin{eqnarray}
\mathcal{F}_1 &=& \frac{W - m_N}{8 \pi W} \sqrt{E_p + m_N} \sqrt{
 E_{p'} + m_N} \left( A_1 + (W - m_N) A_4 - \frac{2 m_N \nu_B}{W - m_N} (A_3 - A_4) + 
   \frac{Q^2}{W - m_N} A_6 \right) ,\\
\mathcal{F}_2 &=& \frac{W+m_N}{8 \pi W} |\vec{q}| \sqrt{\frac{E_p - m_N}{E_{p'}+m_N}} \left( - A_1 + (W+m_N)A_4 - \frac{2 m_N \nu_B}{W+m_N}(A_3-A_4) + \frac{Q^2}{W+m_N} A_6 \right),  \\
\mathcal{F}_3 &=& \frac{W+m_N}{8 \pi W} |\vec{q}| \sqrt{E_p - m_N} \sqrt{
 E_{p'} + m_N} \left( \frac{2 W^2 - 2 m_N^2 + Q^2}{2(W+m_N)} A_2 + A_3 - A_4 - \frac{Q^2}{W+m_N} A_5 \right) ,\\
\mathcal{F}_4 &=& \frac{W-m_N}{8 \pi W} |\vec{q}|^2 \sqrt{\frac{E_p + m_N}{E_{p'}+m_N}} \left( - \frac{2 W^2 - 2 m_N^2 + Q^2}{2(W-m_N)} A_2 + A_3 - A_4 + \frac{Q^2}{W-m_N} A_5 \right), \\
\mathcal{F}_5 &=& \frac{E_\gamma}{8 \pi W} \sqrt{\frac{E_{p'}+m_N}{E_p + m_N}} \mbox{\Huge{ $\lbrace$ }} [E_p + m_N]A_1 \nn \\
&& + \left[ 4 m_N \nu_B \left( W -\frac{3}{4} E_\gamma \right) - |\vec{p}_\gamma |^2 W + E_\pi \left( W^2 - m_N^2 + \frac{1}{2} Q^2 \right) \right] A_2 \nn \\
&& + [ E_\pi(W+m_N) + 2m_N \nu_B ] A_3 \nn \\
&&+ [(E_p+m_N)(W-m_N) - E_\pi (W+m_N) - 2m_N \nu_B]A_4 \nn \\
&& + \left[ 2m_N \nu_B E_\gamma - E_\pi Q^2 \right] A_5 - [(E_p + m_N)(W-m_N)]A_6 \mbox{\Huge{ $\rbrace$ }} ,\\
\mathcal{F}_6 &=&  \frac{E_\gamma}{8 \pi W} \frac{|\vec{q}|}{\sqrt{(E_{p'}+m_N)(E_p-m_N)}} \mbox{\Huge{ $\lbrace$ }} -[Ep-m_N]A_1 \nn \\
&& + \left[ |\vec{p}_\gamma |^2 W - 4 m_N \nu_B \left( W - \frac{3}{4} E_\gamma \right) - E_\pi \left( W^2 - m_N^2 + \frac{1}{2} Q^2 \right) \right] A_2 \nn \\
&& + \left[ E_\pi (W-m_N) + 2 m_N \nu_B \right] A_3 \nn \\
&& + \left[ (E_p - m_N)(W+m_N) - E_\pi (W-m_N) -2 m_N \nu_B \right] A_4 \nn \\
&& + \left[ E_\pi Q^2 - 2 m_N \nu_B E_\gamma \right] A_5 - [(E_p - m_N)(W+m_N)] A_6 \mbox{\Huge{ $\rbrace$ }},
\label{eq:F6}
\end{eqnarray}
where $\nu_B=- \dfrac{k \cdot q}{2 m_N} = - \dfrac{s+u - 2m_N^2}{4 m_N}$. Some care is needed here because different conventions for these functions can be found in the literature\footnote{ For instance, in Ref.~\cite{Dennery:1961zz}, the expressions for $\mathcal{F}_5$ and $\mathcal{F}_6$ are quite different from ours, {\it i.e.}, $\mathcal{F}_5=\mathcal{F}_5^{\mbox{\cite{Dennery:1961zz}}} + \mathcal{F}_1 +\cos \theta_\pi \mathcal{F}_3 $ and $\mathcal{F}_6=  \mathcal{F}_6^{\mbox{\cite{Dennery:1961zz}}}+ \cos \theta_\pi \mathcal{F}_4$. For the rest of the amplitudes, $\mathcal{F}_1,\cdots , \mathcal{F}_4$ there are only global factors in the comparison.}.

\subsection{Amplitude pieces}
\label{app:amplitudes}

\subsubsection{$\mathcal{O}(q^1)$ order}

\begin{align}
\mathcal{M}^{\mu \; (1)}_{(a)} &= C_{I}^{(1)} \frac{e g }{F} V_1^{\mu },\\
\mathcal{M}^{\mu \; (1)}_{(b)} &= C_{II}^{(1)} \frac{e g}{F} \left(\frac{\left(m_N^2-s\right) V_1^{\mu }}{s-m_2^2}-\frac{\left(m_N+m_2\right) \left(2 V_2^{\mu }+V_3^{\mu }+V_4^{\mu }-V_5^{\mu }\right)}{m_2^2-s}\right),\\
\mathcal{M}^{\mu \; (1)}_{(c)} &= C_{III}^{(1)} \frac{e g}{F} \left(\frac{\left(u-m_N^2\right) V_1^{\mu }}{u-m_2^2}-\frac{\left(m_N+m_2\right) \left(2 V_2^{\mu }-V_3^{\mu }+V_4^{\mu }-V_5^{\mu }\right)}{m_2^2-u}\right),\\
\mathcal{M}^{\mu \; (1)}_{(d)} &= C_{IV}^{(1)} \frac{\sqrt{2} e m_N g \left(2 V_3^{\mu }-V_4^{\mu }\right)}{F \left(-2 m_N^2+Q^2+s+u\right)}.
\end{align}
The constants $C_I^{(1)}, \dots , C_{IV}^{(1)}$ are given in Table \ref{tab:coefs1} for each reaction channel. The amplitudes $\mathcal{M}^{\mu \; (1)}_{(b)}$ and $\mathcal{M}^{\mu \; (1)}_{(c)}$ are actually a combination of $\mathcal{O}(q^1)$ and $\mathcal{O}(q^2)$ orders due to the insertion of the nucleon mass at $\mathcal{O}(q^2)$, $m_2$, in the $N$-propagator. This automatically generates the above diagrams at $\mathcal{O}(q^1)$ with the chiral nucleon mass, $m$, in the propagator and the diagrams at $\mathcal{O}(q^2)$ with the insertion of a vertex proportional to $c_1$ in the $N$-propagator, plus higher order small terms. As always, for the external legs we use physical masses.

\begin{table}[H] \centering
\begin{tabular}
[c]{|l|c|c|c|c|}\hline
Channel & $C_{I}^{(1)}$ & $C_{II}^{(1)}$ & $C_{III}^{(1)}$ & $C_{IV}^{(1)}$ \\ \hline
$\gamma^* p \rightarrow p \pi^0$ & $0$                   & $\frac{1}{2}$        & $\frac{1}{2}$ & $0$ \\
$\gamma^* p \rightarrow n \pi^+$ & $\frac{1}{\sqrt{2}}$  & $\frac{1}{\sqrt{2}}$ & $0$ & $-1$ \\
$\gamma^* n \rightarrow p \pi^-$ & $-\frac{1}{\sqrt{2}}$ & $0$                  & $\frac{1}{\sqrt{2}}$ & $1$ \\
$\gamma^* n \rightarrow n \pi^0$ & $0$                   & $0$                  & $0$ & $0$ \\ \hline
\end{tabular}
\caption{Tree level amplitude constants for each channel at $\mathcal{O}(q^1)$.\label{tab:coefs1}}
\end{table}

\subsubsection{$\mathcal{O}(q^2)$ order}

\begin{align}
\mathcal{M}^{\mu \; (2)}_{(b)} &= C_{II}^{(2)} \frac{e g_A}{F_\pi} \left(-\frac{2 V_6^{\mu }+V_7^{\mu }}{m_N^2-s}-\frac{\left(3 m_N^2+s\right) \left(V_4^{\mu }-V_5^{\mu }\right)}{2m_N( m_N^2- s)}-V_1^{\mu }\right) ,\\
\mathcal{M}^{\mu \; (2)}_{(c)} &= C_{III}^{(2)} \frac{e g_A}{F_\pi} \left(\frac{V_7^{\mu }-2 V_6^{\mu }}{m_N^2-u}-\frac{\left(3 m_N^2+u\right) (V_4^{\mu }-V_5^{\mu })}{2m_N( m_N^2- u)}+V_1^{\mu }\right).
\end{align}
The constants $C_{II}^{(2)}$ and $C_{III}^{(2)}$ are given in Table~\ref{tab:coefs2}.

\begin{table}[H] \centering
\begin{tabular}
[c]{|l|c|c|c|c|}\hline
Channel & $C_{I}^{(2)}$ & $C_{II}^{(2)}$ & $C_{III}^{(2)}$ & $C_{IV}^{(2)}$ \\ \hline
$\gamma p \rightarrow p \pi^0$&$0$ & $\frac{1}{2}(c_6 + c_7)$& $\frac{1}{2}(c_6 + c_7)$ & $0$ \\
$\gamma p \rightarrow n \pi^+$&$0$ & $\frac{1}{\sqrt{2}}(c_6 + c_7)$& $\frac{1}{\sqrt{2}}c_7$ & $0$  \\
$\gamma n \rightarrow p \pi^-$&$0$ & $\frac{1}{\sqrt{2}}c_7$ & $\frac{1}{\sqrt{2}}(c_6 + c_7)$ & $0$ \\
$\gamma n \rightarrow n \pi^0$&$0$ & $-\frac{1}{2}c_7$       & $-\frac{1}{2}c_7$          & $0$ \\ \hline
\end{tabular}
\caption{Tree level amplitude  constants for each channel at $\mathcal{O}(q^2)$.\label{tab:coefs2}}
\end{table}

\subsubsection{$\mathcal{O}(q^{5/2})$ order}

\begin{equation}
\begin{split}
 \mathcal{M}^{\mu \; (5/2)}_{(b)} =  D_{II} &  \frac{e h_A g_M}{24 F_\pi m_N m_{\Delta } \left(m_{\Delta }+m_N\right) \left(m_{\Delta }^2-s-i \Gamma _{\Delta } m_{\Delta }\right)} \Biggl[  \\
& \left\{  m_N \left(m_N^4+m_N^2 \left(-M_\pi^2+Q^2+2 s\right)+M_\pi^2 \left(s-Q^2\right)+s \left(-Q^2+3 s-6 u\right)\right) \right.  \\
& \left. -m_{\Delta } \left(3 m_N^4+m_N^2 \left(M_\pi^2-Q^2-10 s\right)+M_\pi^2 \left(5 Q^2-s\right)+s \left(Q^2+s+6 u\right)\right)\right\} V_1^{\mu}  \\
+ & \left\{ 2 Q^2 \left(m_N^2-4 m_N m_{\Delta }-M_\pi^2-5 s\right)+6 \left(m_N m_{\Delta }+s\right) \left(2 m_N^2-s-u\right) \right\} V_2^{\mu}  \\
+ & \left\{ Q^2 \left(m_N^2-10 m_N m_{\Delta }-M_\pi^2+s\right)+3 (s-u) \left(m_N m_{\Delta }+s\right)  \right\} V_3^{\mu}  \\
+ & \left\{ Q^2 \left(m_N^2-4 m_N m_{\Delta }-M_\pi^2-5 s\right)+m_N m_{\Delta } \left(2 m_N^2+4 M_\pi^2-3 (5 s+u)\right)\right.  \\
& \left. -s \left(6 m_N^2-4 M_\pi^2+7 s+3 u\right)  \right\} V_4^{\mu}  \\
+ & \left\{ -m_N^4+m_{\Delta } \left(8 m_N^3-4 m_N M_\pi^2+8 m_N s\right)+Q^2 \left(-m_N^2+4 m_N m_{\Delta }+M_\pi^2+5 s\right) \right.  \\
& \left. +m_N^2 \left(M_\pi^2+6 s\right)+s \left(-5 M_\pi^2+5 s+6 u\right)  \right\} V_5^{\mu} \\
+ & \left\{ 2 m_N \left(m_N^2-M_\pi^2-9 s\right)-2 m_{\Delta } \left(9 m_N^2+M_\pi^2+2 s-3 u\right)  \right\} V_6^{\mu}  \\
+ & \left\{  -m_{\Delta } \left(3 m_N^2+M_\pi^2-4 s-3 u\right)+m_N \left(m_N^2-M_\pi^2+3 s\right)+6 Q^2 m_{\Delta }  \right\} V_7^{\mu}  \\
+ & \left\{  m_{\Delta } \left(-5 m_N^2-5 M_\pi^2+2 s+3 u\right)+m_N \left(m_N^2-M_\pi^2-s\right)  \right\} V_8^{\mu} \Biggr],
\end{split}
\end{equation}

\begin{equation}
\begin{split}
 \mathcal{M}^{\mu \; (5/2)}_{(c)} = D_{III} & \frac{e h_A g_M}{24 F_\pi m_N m_{\Delta } \left(m_{\Delta }+m_N\right) \left(m_{\Delta }^2-u\right)} \Biggl[  \\
& \left\{ m_N \left(u \left(2 m_N^2+M_\pi^2-Q^2-6 s\right)+(m_N-M_\pi) (m_N+M_\pi) \left(m_N^2+Q^2\right)+3 u^2\right) \right. \\
& \left. -m_{\Delta } \left(3 m_N^4+m_N^2 \left(M_\pi^2-Q^2-10 u\right)+M_\pi^2 \left(5 Q^2-u\right)+u \left(Q^2+6 s+u\right)\right) \right\} V_1^{\mu } \\
+ & \left\{ 2 Q^2 \left(-m_N^2+4 m_N m_{\Delta }+M_\pi^2+5 u\right)-6 \left(2 m_N^2-s-u\right) \left(m_N m_{\Delta }+u\right) \right\} V_2^{\mu } \\
+ & \left\{ Q^2 \left(m_N^2-10 m_N m_{\Delta }-M_\pi^2+u\right)-3 (s-u) \left(m_N m_{\Delta }+u\right) \right\} V_3^{\mu } \\
+ & \left\{ -2 m_N^4+Q^2 \left(-m_N^2+4 m_N m_{\Delta }+M_\pi^2+5 u\right) \right. \\
& \left. +m_N m_{\Delta } \left(18 m_N^2-4 M_\pi^2-3 s+u\right)+2 m_N^2 \left(M_\pi^2+3 u\right)+3 u \left(-2 M_\pi^2+3 s+u\right) \right\} V_4^{\mu } \\
+ & \left\{ m_N^4+Q^2 \left(m_N^2-4 m_N m_{\Delta }-M_\pi^2-5 u\right)+4 m_N m_{\Delta } \left(-2 m_N^2+M_\pi^2-2 u\right) \right.  \\
& \left. -m_N^2 \left(M_\pi^2+6 u\right)+u \left(5 M_\pi^2-6 s-5 u\right) \right\} V_5^{\mu } \\
+ & \left\{ 2 m_{\Delta } \left(9 m_N^2+M_\pi^2-3 s+2 u\right)+ 2 m_N \left(-m_N^2+M_\pi^2+9 u\right) \right\} V_6^{\mu } \\
+ & \left\{ -m_{\Delta } \left(3 m_N^2+M_\pi^2-3 s-4 u\right)+m_N \left(m_N^2-M_\pi^2+3 u\right)+6 Q^2 m_{\Delta } \right\} V_7^{\mu } \\
+ & \left\{ m_{\Delta } \left(-5 m_N^2-5 M_\pi^2+3 s+2 u\right)+m_N \left(m_N^2-M_\pi^2-u\right) \right\} V_8^{\mu } \Biggr],
\end{split}
\end{equation}
where $\Gamma_{\Delta}(s)$ is the energy-dependent width given by \cite{Gegelia:2016pjm}
\begin{align}
    \Gamma_{\Delta}(s) = \frac{(h_A/2)^2\Lambda^{3/2}(s, M_\pi^2,m_N^2)}{192 \pi F_\pi s^3} \left( (s-M_\pi^2 + m_N^2) m_\Delta +2 s m_N\right)\theta \left(s - (m_N+M_\pi)^2 \right),
\end{align}
with $\Lambda(x,y,z)=(x-y-z)^2-4yz$ the Källén function and $\theta(x)$ the unit step function. The constants $D_{II}$ and $D_{III}$ are presented in Table~\ref{tab:coefs25}.

\begin{table}[H] \centering
\begin{tabular}
[c]{|l|c|c|}\hline
Channel & $D_{II}$ & $D_{III}$ \\ \hline
$\gamma^* p \rightarrow p \pi^0$ & $1$& $-1$  \\
$\gamma^* p \rightarrow n \pi^+$ & $-\frac{1}{\sqrt{2}}$& $-\frac{1}{\sqrt{2}}$   \\
$\gamma^* n \rightarrow p \pi^-$ & $\frac{1}{\sqrt{2}}$ & $\frac{1}{\sqrt{2}}$  \\
$\gamma^* n \rightarrow n \pi^0$ & $1$ & $-1$  \\ \hline
\end{tabular}
\caption{Tree level amplitude  constants for each channel at $\mathcal{O}(q^{5/2})$. \label{tab:coefs25}}
\end{table}

\subsubsection{$\mathcal{O}(q^3)$ order}

\begin{align}
\mathcal{M}^{\mu \; (3)}_{(a)}= & C_{Ia}^{(3)} \frac{e}{F_\pi m_N} \mbox{\Huge $ ( $} V_4^{\mu } \left(-4 m_N^2+2 M_\pi^2-2 Q^2-3 s-u\right)+2 V_5^{\mu } \left(2 m_N^2-M_\pi^2+Q^2+s+u\right) \nn \\
&  +V_2^{\mu } \left(4 m_N^2-2 \left(2 Q^2+s+u\right)\right)+2 m_N (s-u) V_1^{\mu }-8 m_N V_6^{\mu }+(s-u) V_3^{\mu } \mbox{\Huge $ ) $} \nn \\
 + & C_{Ib}^{(3)} \left[ \frac{\sqrt{2} \left(d_{18}-2 d_{16}\right) e M_\pi^2 }{F_\pi}V_1^{\mu } \right. \nn \\
&+\frac{d_{20} e}{\sqrt{2} F_\pi m_N^2} \left(\frac{1}{4} (V_7^{\mu }-V_8^{\mu }) \left(2 m_N^2+2 M_\pi^2-s-u\right) +\frac{1}{2} (s-u) V_6^{\mu } \right. \nn \\
&  \left. +\frac{1}{4} V_1^{\mu } \left( 2 m_N^4+Q^2 \left(-2 m_N^2-2 M_\pi^2+s+u\right)+2 m_N^2 \left(M_\pi^2-s-u\right)-M_\pi^2 (s+u)+2 s u\right) \right) \nn \\
& \left. + \frac{d_{21} e}{F_\pi} \left(\frac{ \left(2 m_N^2-s-u\right)}{\sqrt{2}}V_1^{\mu } +\sqrt{2} V_7^{\mu }\right) + \frac{d_{22} e }{F_\pi}  \left(\frac{V_1^{\mu } \left(-2 m_N^2+2 Q^2+s+u\right)}{2 \sqrt{2}}+\frac{V_8^{\mu }-V_7^{\mu }}{\sqrt{2}}\right) \right] , \\
\mathcal{M}^{\mu \; (3)}_{(b.\gamma 1 \pi 3)} =& C_{IIa}^{(3)} \frac{\left(d_{18}-2 d_{16}\right) e M_\pi^2 }{F_\pi} \left( V_1^{\mu } + \frac{2 m_N \left(2 V_2^{\mu }+V_3^{\mu }+V_4^{\mu }-V_5^{\mu }\right)}{m_N^2-s}\right), \\
\mathcal{M}^{\mu \; (3)}_{(b.\gamma 3 \pi 1)} =& C_{IIb}^{(3)} \frac{e g_A}{4 F_\pi} \left(Q^2 \left(\frac{2 \left(2 V_6^{\mu }+V_7^{\mu }\right)}{m_N^2-s}-\frac{\left(3 m_N^2+s\right) \left(2 V_2^{\mu }+V_3^{\mu }\right)}{m_N(m_N^2- s)}\right)+\left(\frac{s}{m_N}+3 m_N\right) V_4^{\mu }-2 V_8^{\mu }\right), \\
\mathcal{M}^{\mu \; (3)}_{(c.\gamma 1 \pi 3)} =& C_{IIIa}^{(3)} \frac{\left(d_{18}-2 d_{16}\right) e M_\pi^2 }{F_\pi} \left(- V_1^{\mu } + \frac{2 m_N \left(2 V_2^{\mu }-V_3^{\mu }+V_4^{\mu }-V_5^{\mu }\right)}{m_N^2-u}\right), \\
\mathcal{M}^{\mu \; (3)}_{(c.\gamma 3 \pi 1)} =& C_{IIIb}^{(3)} \frac{e g_A}{4 F_\pi} \left(Q^2 \left(\frac{2 (2 V_6^{\mu }-V_7^{\mu})}{m_N^2-u}-\frac{\left(3 m_N^2+u\right) \left(2 V_2^{\mu }-V_3^{\mu }\right)}{m_N(m_N^2- u)}\right)-\left(\frac{u}{m_N}+3 m_N\right) V_4^{\mu }+2 V_8^{\mu }\right), \\
\mathcal{M}^{\mu \; (3)}_{(d:N3 \pi 2)} =& C_{IV}^{(3)} \frac{2 \sqrt{2} \left(d_{18}- 2 d_{16}\right) e m_N M_\pi^2 \left(2 V_3^{\mu }-V_4^{\mu }\right)}{F_\pi \left(-2 m_N^2+Q^2+s+u\right)}, \\
\mathcal{M}^{\mu \; (3)}_{(d:N1 \pi 4)} =& C_{IV}^{(3)} \frac{e g_A}{F_\pi^3} \left(-\frac{2 \sqrt{2} l_4 m_N M_\pi^2 \left(2 V_3^{\mu }-V_4^{\mu }\right)}{-2 m_N^2+Q^2+s+u}-\frac{\sqrt{2} l_6 m_N \left(V_4^{\mu } \left(-2 m_N^2+s+u\right)+2 Q^2 V_3^{\mu }\right)}{-2 m_N^2+Q^2+s+u}\right), \\
\mathcal{M}^{\mu \; (3)}_{(g)} = & C_{IV}^{(1)} \frac{\sqrt{2} e m_N g_A \left(2 V_3^{\mu }-V_4^{\mu }\right)}{F_\pi \left(-2 m_N^2+Q^2+s+u\right)} \xi,
\end{align}
where 
\begin{equation}
\begin{split}
\xi = &  \frac{2 M_\pi^2}{F_\pi} \left( \frac{M_\pi^2}{2m_N^2-Q^2-s-u} l_3 - l_4 \right),
\end{split}
\end{equation}
 and the corresponding constants $C_{Ia}^{(3)}, \dots , C_{IV}^{(3)}$ are defined in Table~\ref{tab:coefs3}.

\begin{table}[H] \centering
\begin{tabular}
[c]{|l|c|c|c|c|c|c|c|}\hline
Channel &$C_{Ia}^{(3)}$ & $C_{Ib}^{(3)}$ & $C_{IIa}^{(3)}$& $C_{IIb}^{(3)}$ & $C_{IIIa}^{(3)}$ & $C_{IIIb}^{(3)}$ & $C_{IV}^{(3)}$ \\ \hline
$\gamma^* p \rightarrow p \pi^0$&$d_8+d_9$     & $0$ & $1$       & $2d_7+d_6$          & $1$       & $2d_7+d_6$ & $0$ \\
$\gamma^* p \rightarrow n \pi^+$&$\sqrt{2}d_9$ & $-1$& $\sqrt{2}$& $\sqrt{2}(2d_7+d_6)$& $0$       & $\sqrt{2}(2d_7-d_6)$&$1$ \\
$\gamma^* n \rightarrow p \pi^-$&$\sqrt{2}d_9$ & $1$ & $0$       & $\sqrt{2}(2d_7-d_6)$& $\sqrt{2}$& $\sqrt{2}(2d_7+d_6)$&$-1$ \\
$\gamma^* n \rightarrow n \pi^0$&$d_8-d_9$     & $0$ & $0$       & $-(2d_7-d_6)$       & $0$       & $-(2d_7-d_6)$&$0$ \\ \hline
\end{tabular}
\caption{Tree level amplitude  constants for each channel at $\mathcal{O}(q^3)$. \label{tab:coefs3}}
\end{table}

\subsection{EOMS $\beta$ functions}
\label{app:eomsbeta}

For the parameters $m$ and $g$, from $\mathcal{L}_{\pi N}^{(1)}$, we get
\begin{align}
 \widetilde\beta_{m} =  - \frac{3}{2} g^2 \overline{A}_0\left[m^2\right], \qquad
 \widetilde\beta_g = g^3 m + \frac{\left(2-g^2\right) g }{m} \overline{A}_0\left[m^2\right],
\label{eq:eomsgm}
\end{align}
where
\begin{equation}
\overline{A}_0[m^2]    = -m^2 \log\frac{m^2}{\mu^2},
\end{equation}
is the ${\widetilde{\rm MS}}$-renormalized scalar 1-point Passarino-Veltman function with $\mu$ the renormalization scale introduced in the dimensional regularization.
For the second order LECs in $\mathcal{L}_N^{(2)}$ we 
have~\cite{Fuchs:2003ir}
\begin{align}
\widetilde \beta_{c_1} = \frac{3}{8} g^2 + \frac{3g^2}{8m^2} \overline{A}_0[m^2], \qquad
\widetilde \beta_{c_6} = - 5 g^2 m, \qquad  
\widetilde \beta_{c_7} = 4 g^2 m .
\label{eq:eomsc}
\end{align}
In this case, as we are using in practice the $\mathcal{O}(p^2)$ nucleon mass, $m_2=m-4c_1M_\pi^2$, it's easy to see that the corresponding EOMS shift results in
\begin{align}
m_2 =  \widetilde m_2 + \frac{m \left( \widetilde \beta_m - 4 M_\pi^2 \widetilde \beta_{c_1} \right)}{16 \pi^2 F^2}.
\end{align}

\subsection{Wave function renormalization}\label{app:wfr}
The wave function renormalization of the external legs, in the EOMS scheme, is written as
\begin{align}
\mathcal{Z}_N= &1 + \delta^{(2)}_{\mathcal{Z}_N} + \mathcal{O}(p^3), \qquad \mathcal{Z}^{(2)}_\pi=1 + \delta^{(2)}_{\mathcal{Z}_\pi} + \mathcal{O}(p^3),\\
\intertext{ where}
\delta^{(2)}_{\mathcal{Z}_N} = &-\frac{3 g_A^2}{64 \pi ^2 F_\pi^2 \left(M_\pi^2-4 m_N^2\right)} \Bigg \{ 4 M_\pi^2 \left(A_0\left[m_N^2\right]+\left(M_\pi^2-3 m_N^2\right) B_0\left[m_N^2,M_\pi^2,m_N^2\right]-m_N^2\right) \nn \\
&+ \left(12 m_N^2-5 M_\pi^2\right) A_0\left[M_\pi^2\right] \Bigg \}, \\
\delta^{(2)}_{\mathcal{Z}_\pi} = & -\frac{2}{3 F_\pi^2} \left\lbrace 3 l_4 M_\pi^2 + \frac{A_0\left[M_\pi^2\right]}{16 \pi ^2 }\right\rbrace .
\end{align}

\subsection{Chiral expansions for physical quantities in the EOMS scheme}\label{app:chiralexpansions}

For the nucleon mass, $m_N$, we have 
\begin{align}
m_N=&\widetilde{m} -4 \widetilde{c}_1  M_\pi^2 + \widetilde{\delta}^{(3)}_{m} + \mathcal{O}\left( p^4 \right),
\label{eq:m3}\\
\widetilde m_2 = & \widetilde m -4 \widetilde c_1 M_\pi^2 = m_N - \widetilde\delta^{(3)}_{m}+ \mathcal{O}\left( p^4 \right),
\label{eq:m2}\\
\intertext{with}
\widetilde{\delta}^{(3)}_{m}=& \frac{3 g_A^2 m_N \, M_\pi^2}{32 \pi ^2 F_\pi^2} \left\lbrace \overline{B}_0\left[m_N^2,M_\pi^2,m_N^2\right] -\left(1+\frac{\overline{A}_0\left[m_N^2\right]}{m_N^2} \right) \right\rbrace.
\end{align}

For the pion mass we have
\begin{align}
M_\pi^2 =& M^2\left( 1+ \delta^{(2)}_{M_\pi}\right) + \mathcal{O}\left( p^6 \right), \\
\intertext{where}
\delta^{(2)}_{M_\pi} = & \frac{2 l_3^r M_\pi^2}{F_\pi^2} - \frac{\overline{A}_0[M_\pi^2]}{32 \pi^2 F_\pi^2}.
\end{align}
For the axial coupling constant, we have 
\begin{align}
g_A = &\widetilde g \left( 1 + \frac{ 4 d^r_{16} M_\pi^2}{\widetilde g} + \widetilde \delta^{(2)}_{g_A}  + \mathcal{O}(p^3) \right) \\
\intertext{where}
\widetilde \delta^{(2)}_{g_A} =& \frac{1}{16 \pi ^2 F_\pi^2 \left(4 m_N^2-M_\pi^2\right)} \Bigg \{ 4 g_A^2 M_\pi^2 \overline{A}_0\left[m_N^2\right]+\left(\left(8 g_A^2+4\right) m_N^2-\left(4 g_A^2+1\right) M_\pi^2\right) \overline{A}_0\left[M_\pi^2\right] \nonumber \\
& + M_\pi^2 \left(\left(\left(3 g_A^2+2\right) M_\pi^2-8 \left(g_A^2+1\right) m_N^2\right) \overline{B}_0\left[m_N^2,M_\pi^2,m_N^2\right]-4 g_A^2 m_N^2\right) \Bigg \},
\end{align}
For the pion decay constant
\begin{align}
 F_\pi =& F \left( 1+ \delta^{(2)}_{F_\pi}  + \mathcal{O}(p^3) \right),\\
\intertext{where}
\delta^{(2)}_{F_\pi} =& \frac{l_4^r M_\pi^2}{F_\pi^2} + \frac{\overline{A}_0\left[M_\pi^2\right]}{16 \pi ^2 F_\pi^2}.
\end{align}
Note here that $l_4^r$ and $d_{16}^r$ are $\widetilde{\rm MS}$-renormalized LECs.

\bibliography{references}

\end{document}